\documentclass[lettersize,journal]{IEEEtran}

\usepackage{alltt}
\usepackage{relsize}
\usepackage{booktabs}
\usepackage{mdframed}
\usepackage{color}
\usepackage{amsfonts}
\usepackage{amssymb}
\usepackage{boxedminipage}
\usepackage{fancyvrb}

\usepackage{url}
\usepackage{fancybox}%
\usepackage{placeins}
\usepackage{adjustbox}
 \usepackage{verbatim}
\usepackage{lscape}

\definecolor{mygray}{rgb}{0.7,0.7,0.7}

{\begin{list}{}%
         {\setlength{\leftmargin}{#1}}%
         \item[]%
}
{\end{list}}

\newcommand{\conclusionbox}[1]{%
	\vspace{1.9mm}
  \noindent
	\framebox[\columnwidth][c]{%
		\parbox[b]{0.97\columnwidth}{%
			{\em #1}
		}
	}
	\vspace{-4.0mm}
}

\newcommand{\smalltt}[1]{\ifmmode{\mbox{\smaller\texttt{#1}}}\else{\smaller\tt #1}\fi}

{ \medskip
  \noindent
  \let\emph=\textbf
  \begin{boxedminipage}{\columnwidth}\em}%
{ \end{boxedminipage}}

\newdimen\qdx
\newdimen\qda
\newdimen\qdb
\def\rrrr#1#2#3#4{\newdimen\qd\qd=#4 %
\qdx=\qd\multiply\qdx by 5\divide\qdx by 4
\qda=\qd
\qdb=\qd
\multiply\qda by #1\divide\qda by #3\multiply\qdb by #2\divide\qdb by #3\advance\qdb by -\qda
    \leavevmode\hbox to \qdx{\hfil\vbox{%
    \hbox{\vrule\vbox{\hrule\hbox to 1\qd
            {\vrule depth0pt height0.7ex width \qda\color{mygray}%
 \vrule depth0pt height0.7ex width \qdb\hfill}\hrule}\vrule}
    }\hfil}}

  \renewenvironment{thebibliography}[1]{%
    \begin{oldthebibliography}{#1}%
       \setlength{\parskip}{0.2ex}%
       \setlength{\itemsep}{0.2ex}%
  }%
  {%
    \end{oldthebibliography}%
  }

\usepackage[caption=false,font=normalsize,labelfont=sf,textfont=sf]{subfig}
\usepackage{graphicx}
\usepackage[shortlabels,inline]{enumitem}
\setenumerate[1]{leftmargin=1.0cm} 

\usepackage{graphicx}
\usepackage{amsmath}
\usepackage{multirow}
\usepackage[english]{babel}
\usepackage{threeparttable}

\usepackage{cite}

\usepackage{balance}
\usepackage{makecell}
\usepackage{enumitem}

\begin{document}

\author{
Mohamed Sami Rakha,
Andriy Miranskyy \IEEEmembership{Member, IEEE},
Daniel Alencar da Costa

\thanks{
 Mohamed Sami Rakha and Andriy Miranskyy are with the Department of Computer Science,
Toronto Metropolitan University, Canada. 
E-mails:  \{rakha,avm\}@torontomu.ca.
}
\thanks{Daniel da Costa is with the School of Computing, University of Otago, New Zealand. 
E-mail: danielcalencar@otago.ac.nz.
}
}

\def\RQOne{\textbf{Does the performance impact led by hyperparameter tuning vary across the SDP scenarios?
}}

\def\RQTwo{\textbf{How does the performance impact vary for the applied ML algorithms?  }}

\def\RQThree{\textbf{How does the impact of the hyperparameter tuning change when considering different software dataset sizes? }}

\title{Contrasting the Hyperparameter Tuning Impact Across Software Defect Prediction Scenarios}

 \maketitle

\begin{abstract}
Software defect prediction (SDP) is crucial for delivering high-quality software products. The SDP activities help software teams better utilize their software quality assurance efforts, improving the quality of the final product.
Recent research has indicated that prediction performance improvements in SDP are achievable by applying hyperparameter tuning to a particular SDP scenario (e.g., predicting defects for a future version). However, the positive impact resulting from the hyperparameter tuning step may differ based on the targeted SDP scenario. Comparing the impact of hyperparameter tuning across two SDP scenarios is necessary to provide comprehensive insights and enhance the robustness, generalizability, and, eventually, the practicality of SDP modeling for quality assurance. 

Therefore, in this study, we contrast the impact of hyperparameter tuning across two pivotal and consecutive SDP scenarios: (1) Inner Version Defect Prediction (IVDP) and (2) Cross Version Defect Prediction (CVDP). The main distinctions between the two scenarios lie in the scope of defect prediction and the selected evaluation setups. This study's experiments use common evaluation setups, 28 machine learning (ML) algorithms, 53 post-release software datasets, two tuning algorithms, and five optimization metrics. We apply statistical analytics to compare the SDP performance impact differences by investigating the overall impact, the single ML algorithm impact, and variations across different software dataset sizes. 

The results indicate that the SDP gains within the IVDP scenario are significantly larger than those within the CVDP scenario. The results reveal that asserting performance gains for up to 24 out of 28 ML algorithms may not hold across multiple SDP scenarios. Furthermore, we found that small software datasets are more susceptible to larger differences in performance impacts. Overall, the study findings recommend software engineering researchers and practitioners to consider the effect of the selected SDP scenario when expecting performance gains from hyperparameter tuning.

\end{abstract}


\section{Introduction}\label{sec:introduction}

One of the paramount activities in the software engineering (SE) field is software defect prediction (SDP)~\cite{zimmermann2007predicting}. In SDP, an intelligent model is leveraged to predict the defect-prone software modules (e.g., source files) within a software project. If the software defects are known in advance,
the project's quality assurance (QA) team can optimize their efforts, ensuring higher software quality standards. For example, QA personnel can focus their testing efforts on software modules that are predicted to contain more defects.
The optimization of QA resources ultimately reduces project costs, rendering SDP more imperative~\cite{wahono2014neural, bowes2015different}. As such, much research has
invested in models to predict defect-prone software
modules~\cite{zimmermann2007predicting,
d2010extensive, kim2011dealing, he2012investigation,
zhang2014towards,zhang2016cross}. These models are trained using statistical or
Machine Learning (ML) algorithms.

Other domains also use ML to predict defects and failures (e.g., hardware defect prediction), but SDP is unique due to the specificities of the SE field. For example, extracting features from software repositories, such as the product and process metrics~\cite{gong2021revisiting}, is particular to SDP, differing from non-SE prediction problems. Given that software products constantly evolve through SE tasks (e.g., adding new features and refactoring code)~\cite{he2012investigation, zhang2016cross}, SDP must adapt to ever-changing requirements, as software projects could be more susceptible to change than other engineering projects. Consequently, training SDP models involve specific tuning of hyperparameters for ML algorithms. 

Considering the distinct characteristics
of software projects (i.e., size, domain, and number of developers), it is
challenging to determine the best hyperparameter values for a given defect
prediction model. To complicate matters, software defects are diverse in terms of complexity and contextual dependencies, contributing to the challenge of SDP. Unfortunately, researchers frequently use the default settings to train defect
prediction models when performing
research~\cite{tantithamthavorn2016automated}. Relying on default settings in SDP can be detrimental in the long term as they will likely
produce underperforming models. Prior research~\cite{fu2016tuning} has reported that around 80\% of the most cited SDP studies have relied on default settings, resulting in suboptimal model performance.

The positive impact of the hyperparameter tuning for software defect prediction has been widely illustrated by prior research~\cite{wahono2014neural,
tantithamthavorn2016automated, fu2016tuning, osman2017hyperparameter, tantithamthavorn2018impact}.
Wahono et al. ~\cite{wahono2014neural} applied a genetic algorithm to tune
the hyperparameters of a neural network model, including model parameters such as
learning rate, momentum, and training cycles. Tantithamthavorn et
al.~\cite{tantithamthavorn2016automated, tantithamthavorn2018impact} conducted a large study
on the impact of applying various hyperparameter tuning techniques for 26  defect prediction models by
using the \texttt{Caret} API~\cite{XcaretPkg2019}.  Osman et al.~\cite{osman2017hyperparameter} used Multisearch-weka~\cite{WekaSearchPackage} and showed that hyperparameter tuning is a necessary step for software defect prediction. Fu et al.~\cite{fu2016tuning} also studied the impact of hyperparameter tuning on defect prediction models using a differential evolution
technique~\cite{omran2005differential}. Additionally, many recent studies included hyperparameter tuning as an essential step for software defect prediction~\cite{kondo2019impact,li2020understanding,rajbahadur2019impact}.

    Studying the impact of hyperparameter tuning on software defect prediction reveals a diverse landscape of prior research, each using one specific SDP scenario for assessing the impact of hyperparameter tuning on software defect prediction~\cite{wahono2014neural,osman2017hyperparameter,tantithamthavorn2018impact,agrawal2018better,ali2021empirical,fu2016differential,fu2016tuning, lee2022holistic, agrawal2021simpler}. While prior research has demonstrated the benefits of hyperparameter tuning, the performance gains reported are often inconsistent, particularly when comparing results across different SDP scenarios. Some conflicting messages we get from prior research are the ones reported by Tantithamthavorn et al.~\cite{tantithamthavorn2016automated, tantithamthavorn2018impact} and Fu et al.~\cite{fu2016tuning}. While Tantithamthavorn et al. reported large improvements by up to 40\% percentage points and focused on positive gains, Fu et al. results showed more humble improvements and highlighted a possible negative aspect of applying hyperparameter tuning for SDP.

 Motivated by such observations, this study bridges this gap by directly comparing the performance impact of tuning across two pivotal and consecutive SDP scenarios: Inner Version Defect Prediction (IVDP)~\cite{wahono2014neural,osman2017hyperparameter,tantithamthavorn2018impact,agrawal2018better,ali2021empirical} and Cross Version Defect Prediction (CVDP)~\cite{fu2016differential,fu2016tuning, lee2022holistic, agrawal2021simpler} scenarios, offering new insights into when tuning is most beneficial and how it affects model interpretability. These SDP scenarios are closely linked with different phases of the software development life cycle. IVDP, at its core, is centered on predicting defects within a specific software version, whereas CVDP focuses on predicting defects across different software versions. The perceived practicality of each scenario setting depends on factors such as the stage of the software life-cycle, the available data, and the goals of the defect prediction model~\cite{shrikanth2021early}.

\textbf{
This study does not aim to improve SDP performance for a specific scenario. Instead, its primary focus is to better understand the performance impact resulting from applying hyperparameter tuning across SDP scenarios. } In other words, the goal is not to identify the best-performing configuration, but to compare the relative impact of tuning across different evaluation SDP settings, while holding all other factors constant. Both studied SDP scenarios arise from the common objective of improving software quality through defect prediction~\cite{gong2021revisiting}, yet they could represent different stages in the software life cycle. Comparing the impact of hyperparameter tuning across these two scenarios should provide insights that enhance the robustness, generalizability, and practical applicability of SDP modeling in the dynamic landscape of software development. The results of this study are expected to help SE researchers and practitioners understand the differences in SDP performance across two pivotal and consecutive SDP scenarios.

 This study offers a novel large-scale interpretation of how hyperparameter tuning affects SDP in different practical scenarios. To apply this analysis, we statistically compare the impact of hyperparameter tuning by evaluating two tuning algorithms on 28 ML algorithms and 53 post-release software datasets. We explore five optimization metrics: AUC, Recall Rate, Precision, F-measure, and Accuracy. To contrast and highlight the hyperparameter tuning impact variations of the two evaluated SDP scenarios, the study addresses the following research questions\footnote{A preview of the answers is also provided for the reader's convenience.}:

\begin{enumerate}[label=\textbf{RQ\arabic*:}]
	\item {\RQOne} Yes, the IVDP scenario gains larger performance improvements than the CVDP scenario when hyperparameters are tuned for SDP. The difference is statistically significant with non-negligible effect sizes for four optimization metrics.
	\item {\RQTwo} The performance gains of many ML algorithms may not hold up across multiple SDP scenarios.  Up to 22 out of 28 (78\%) applied ML algorithms show statistically significant and non-negligible differences in performance impacts for SDP.

	\item {\RQThree} Small software datasets are affected more than large datasets in terms of varying performance across multiple SDP scenarios.

\end{enumerate}

The  contributions of this study can be summarized as follows: 
\begin{itemize}
\item  \textbf{Comparative Analysis Across SDP Scenarios: } To the best of our knowledge, this study is the first to systematically compare the effects of hyperparameter tuning across distinct SDP scenarios.  This research provides statistical insights into performance differences arising from tuning under diverse software engineering (SE) scenarios, helping SE practitioners understand the variability in tuning impacts across these contexts. 

\item \textbf{Large-Scale and Multi-Dataset Analysis:} This study conducts a large-scale analysis involving approximately  44,520 SDP models using 53 SE post-release datasets and 28 ML algorithms. The scale and diversity of the datasets (i.e., including under-represented datasets) allow the findings to be generalized to various SE contexts, while performance impact is evaluated across five optimization metrics. 

\item \textbf{In-Depth Analysis for Interpretability:} This study delves into the dynamics of hyperparameter tuning and feature importance difference changes across different SDP scenarios. The study presents SDP scenario-specific trends in hyperparameter adjustments, investigates how tuning influences feature importance, and analyzes the role of dataset size, offering fresh insights. This adds to the understanding of how model tuning can alter interpretability in SE applications.

\item  \textbf{Practical Guidelines:} Based on key findings, this study provides actionable recommendations for SE practitioners regarding the choice of ML algorithms and tuning strategies best suited to specific SDP scenarios. These guidelines are grounded in an analysis of IVDP and CVDP results, emphasizing how different ML algorithms respond to tuning under varying data and defect prediction setups.

\end{itemize}

\begin{table*}[!t]
	 
	\centering
	\caption{A list of the studied ML algorithm implementations from the R Caret package~\cite{XcaretPkg2019}. }
	\begin{adjustbox}{scale=0.99}
		\begin{threeparttable}
\begin{tabular}{|c|p{7cm}|p{6.7cm}|}
\hline 
\textbf{Model Family} & \textbf{Model Names (CARET-method value)} & \textbf{Tuned Parameters per Model}\tabularnewline

\hline 
Bayesian & Naive Bayes (naive\_bayes)  & \textbf{naive\_bayes:} (laplace=0, usekernel=FALSE, adjust=0)\tabularnewline
\hline 
Distance Based  & k-Nearest Neighbour (knn), Kernel k-Nearest Neighbors (kknn\dag ) 
  & \textbf{knn:} (k=1), \textbf{kknn:} (kmax=5, distance=2, kernel=optimal)\tabularnewline
\hline 
Regression &  Penalized Multinomial
Regression (multinom)  & \textbf{multinom: }(decay=0)\tabularnewline
\hline 
Neural Network & Neural network (nnet), Averaged Neural network (avNNet),
Multilayer Perceptron (mlp), Voted-Multi-layer Perceptron
(mlpWeightDecay or mlpWD) & \textbf{nnet:} (size=1, decay=0), \textbf{avNNet: }(size=1, decay=0,
bag=FALSE), \textbf{mlp: }(size=1), \textbf{mlpWeightDecay: }(size=1,
decay=0)\tabularnewline
\hline 
Discrimination Analysis & Penalized Discriminant Analysis (pda), Linear Discriminant
Analysis (lda2)& \textbf{pda: }(lambda=1),\textbf{ lda2: }(dimen=1)\tabularnewline
\hline 
Rule-based & Rule-based algorithm (JRip) & \textbf{JRip: }(NumOpt=2, NumFolds=3, MinWeights=2)\tabularnewline
\hline 
Decision Trees- Based & C4.5-like trees (J48), Logistic Model Trees (LMT) 
and Classification And Regression Trees (rpart and rpart2\dag )& \textbf{J48}: (C=0.25, M=2), \textbf{LMT}: (iter=1), \textbf{rpart}:
(cp=0.01), \textbf{rpart2}: (maxdepth=2)\tabularnewline
\hline 
SVM & Support Vector Machines with Linear kernel (SVMRadial), Support
Vector Machines with Radial basis function kernel (SVMLinear)& \textbf{svmRadial: }(sigma=0.5, C=1), \textbf{svmLinear: }(C=1)\tabularnewline
\hline 
Bagging & Random Forest (RF and ranger\dag ), Rotation Forest (rotationForest\dag )
, Regularized Random Forest (RRF\dag ) and Bagged
Adaboost (AdaBag\dag ) & \textbf{RF:} (mtry=10), \textbf{ranger:} (mtry=4, splitrule=gini,
size=1), \textbf{rotationForest}: (K=3, L=10), \textbf{RRF:} (mtry=10,
coefReg=1, coefImp=1), \textbf{AdaBag}: (mfinal=50, maxdepth=1)\tabularnewline
\hline 
Boosting & Gradient Boosting Machine (gbm), Boosted Classification Trees (Ada),  Adaptive Boosting (adaboost.M1), Logistic Regression Boosting (LogitBoost), C5.0
Decision Tree (C5.0)  and Boosted Generalized Linear Model
(glmboost\dag )  & \textbf{gbm}: (n.trees=100, depth=1, shrinkage=1, minobsinnode=10), \textbf{ada}:(iter=50, maxdepth=1, nu=0.1), 
\textbf{adaboost}:(mfinal=50, maxdepth = 1,coeflearn=``Breiman''), \textbf{LogitBoost}:(nIter=11),
\textbf{C5.0}:(trials=1, model=rules, winnow=FALSE), \textbf{glmboost}:
(mstop=100, prune=FALSE)\tabularnewline
\hline 
\end{tabular}
		\begin{tablenotes}
			\item 
				$[\dagger]$ ML algorithms that were not explored in
				previous research about defect prediction
				models
				hyperparamater tuning~\cite{tantithamthavorn2016automated,tantithamthavorn2018impact,fu2016tuning}.
			\item 
			\end{tablenotes}
		\end{threeparttable}

	\end{adjustbox}
 \vspace*{-4mm}
		\label{tab:Tablemodels1}
\end{table*}

\section{Background and Related Work} \label{sec:bg}

In this section, we discuss the related work regarding the impact of hyperparameter tuning on SDP scenarios while also explaining the concepts that will be discussed throughout the paper.

\subsection{Hyperparameter Tuning} \label{subsec:paramSett}
Hyperparameter tuning is the process of choosing the best set of parameter settings for an ML algorithm when solving a certain problem~\cite{duan2003evaluation}. The evaluation of the best parameters is usually guided by a performance measure. In many cases, the performance of an algorithm on a predication task is influenced by its hyperparameter settings. The implementation of an ML algorithm comes with a default hyperparameter setting set by the developers of the ML library. To achieve optimal performance, the practitioner can tune the algorithms' hyperparameters in a process that involves a search space of parameter values and ranges.  Hyperparameter tuning is generally computationally intensive, and the complexity increases with the size of the search space. Currently, there is no empirical evidence identifying which hyperparameters are crucial to tune and which ones yield similar performance when set to reasonable default values~\cite{probst2019tunability}. Table~\ref{tab:Tablemodels1} lists popular ML algorithms applied in this study, along with their hyperparameter default values. 

Machine learning (ML) algorithms such as random forest~(rf) and neural networks~(nnet) require the setting of several hyperparameters that have to
be set before running them. These tuning parameters often have to be carefully
optimized to achieve maximal performance~\cite{eiben2011parameter}. To select the appropriate hyperparameter configuration for a specific software dataset, software defect prediction (SDP) practitioners can use the default values provided by the ML packages (i.e., R Caret package~\cite{XcaretPkg2019}) or manually configure them. Manual configuration can be guided by recommendations from the literature, previous experience, or through trial-and-error. Alternatively, hyperparameter tuning algorithms can be employed. These are data-dependent optimization procedures aimed at minimizing the expected generalization error of the ML algorithm by exploring a hyperparameter search space of potential settings. This process typically involves evaluating predictions on an independent test set or employing a resampling method such as cross-validation and bootstrap~\cite{tantithamthavorn2018impact,kondo2019impact,rajbahadur2019impact, rajbahadur2021impact, jiarpakdee2021practitioners}.

To explore different hyperparameter settings for the SDP scenarios in this study, we define two groups for the hyperparameter settings. First is the  {\em Tuned Settings}, which refers to the parameters generated from the application of a tuning algorithm~\cite{back1996evolutionary}. The tuned settings group in this study includes frequently applied settings, which are  {\em Grid Setting} and  {\em Random Setting}. The grid setting is the results from applying the grid search algorithm~\cite{diaz1983method}, while the random setting is the results from the random search algorithm~\cite{bergstra2012random}. The grid search technique applies a brute-force (or exhaustive) search on a predefined parameter  space~\cite{fu2016differential} while random search randomly samples parameters across a range of parameter space. A {\em parameter space} 
is the space of possible values of a given model parameter, e.g., the number of decision trees~(specified by the ``mtry'' parameter in Table~\ref{tab:Tablemodels1})
in a Random Forest~\cite{ho1995random}. %
The second hyperparameter setting group is the {\em Default Settings}, which refers to the pre-selected parameters by the ML library developers without any tuning or performance evaluation of different hyperparameter settings. We use the settings in this group as the main baseline upon which we demonstrate the impact of the hyperparameter tuning step on defect prediction.

\begin{figure*}[!t]
	\centering
	\vspace*{-5mm}
	\subfloat[\label{fig:Approach1} \small IVDP Experimental Setup ]
 {\includegraphics[width=0.87\linewidth,keepaspectratio]{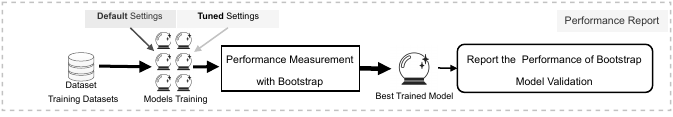}}
	
	\subfloat[\label{fig:Approach2} \small CVDP Experimental Setup ]{\includegraphics[width=0.87\linewidth,keepaspectratio]{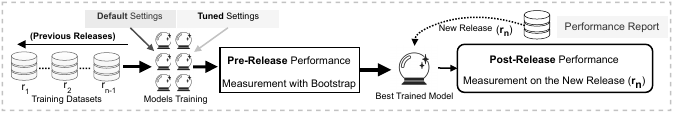}}
	\caption{Overview of the included SDP scenarios experimental setups.}
	\label{fig:ApproachesType}

\captionsetup[subfloat]{font=small}
\vspace*{-2mm}
\end{figure*}
\subsection{SDP Evaluation Scenarios} \label{sec:EvaluationSetup}
In this study, we include two SDP evaluation scenarios frequently used in prior research when hyperparameter tuning is studied or implemented~\cite{fu2016tuning,tantithamthavorn2018impact}. Figure~\ref{fig:Approach1} shows the first scenario evaluation setup, namely IVDP. The IVDP 
is a common setup applied by prior studies on the impact of hyperparameter tuning for defect prediction~\cite{wahono2014neural,osman2017hyperparameter,tantithamthavorn2018impact}. In IVDP, all data instances are used in
the model validation processes, such as 10-fold cross-validation and out-of-sample bootstrap. In this setup, the tuned model is produced by the parameters that lead to the best performance throughout the validation process. In this study, we choose the widely used out-of-sample bootstrap as a validation method. 
Similarly, the performance of the default model is produced by using the
default hyperparameters throughout the bootstrap process. IVDP does not preserve the order of data, such as the release order of a software system.  In the  IVDP, the bootstrap validation is used to train and evaluate the SDP models while exploring various hyperparameter settings.  As illustrated in Figure~\ref{fig:Approach1}, the selected best-tuned model is picked from a list of tuned models.  However, the validation method estimates the performance of defect models while experimenting with different hyperparameters. In this setup, the final tuned model is not tested on unseen data after being selected.

The second Scenario is the CVDP~\cite{fu2016differential,fu2016tuning}. Figure~\ref{fig:Approach2} shows the setup of CVDP in which the input data is split based on the releases. The CVDP Scenario keeps the order of the software releases~\cite{shrikanth2021early}. The previous releases of a software system are used for the normal process of model validation and hyperparameters tuning (as in the IVDP setup, see Figure~\ref{fig:Approach1}), while a new release of a software system is used to evaluate the final tuned model selected out of the validation process. The data of the new software release is unseen testing data for the final tuned model.

\subsection{Related Work} \label{subsec:relatedWork}
This study aims to contrast the impact of hyperparameter tuning across SDP scenarios, exploring generalizability and assessing the differences that should help the researchers refine SDP best practices. Prior studies have investigated the impact of various hyperparameter tuning without considering the role that an SDP scenario selection could play in the conclusions.  Therefore, in this section,  we survey the related research on the impact  of hyperparameter tuning on SDP for each of the studied two major SDP scenarios:

\subsubsection{Inner Version Defect Prediction (IVDP)} \label{subsec:IVDPRelated}
In this SDP scenario, both training and testing data originate from a single project version~\cite{tantithamthavorn2016empirical}.  Typically, a portion of the data from a specific version is used for training a model, and the remaining is used for testing the trained SDP model. Several studies have investigated the impact of hyperparameter tuning in this scenario~\cite{wahono2014neural, tantithamthavorn2018impact, khan2020hyper, nevendra2022empirical}.

Wahono et al$.$~\cite{wahono2014neural} proposed a combination of genetic algorithm (GA) and bagging technique for improving the accuracy of software defect prediction.  More specifically, a genetic algorithm is used to tune
the hyperparameters of a neural network model, including parameters such as
learning rate, momentum, and training cycles. The results from testing on nine NASA (MDP) datasets~\cite{gray2012reflections} reveal statistically significant higher performance in contrast to default neural network models. Another study is by Osman et al$.$~\cite{osman2017hyperparameter} in which the impact of hyper-parameter tuning is investigated using two machine learning algorithms: K-Nearest Neighbors (KNN) and Support Vector Machines (SVM). A hill-climbing grid search algorithm was used to search for the optimal hyperparameter values with experiments on five open-source datasets. The results show prediction accuracy improvements by up to 20\% for KNN and by up to 10\% for SVM.

Tantithamthavorn et al$.$~\cite{tantithamthavorn2016automated} explored the
impact of applying grid optimization to the parameters of software defect
models across numerous datasets and ML algorithms. Tantithamthavorn et al$.$~\cite{tantithamthavorn2016automated} study 18
publicly available datasets, using 26 ML algorithms from the \texttt{Caret} API. The
authors demonstrated that grid optimization can improve the Area Under the receiver operator characteristic Curve (AUC) values of
defect prediction models by up to 40\% percentage points for ML algorithms (as in the case of C5.0 boosting~\cite{defectDataR2019}).  After grid optimization, the results reveal that the likelihood of obtaining a top-performing model is increased by 83\%. For example, Tantithamthavorn et al$.$~\cite{tantithamthavorn2016automated} find that the
likelihood of a C5.0 model type producing a top-performing model is
significantly higher than the likelihood of a Random Forest model type.  In a follow-up study, Tantithamthavorn et al$.$~\cite{tantithamthavorn2018impact} extended prior work~\cite{tantithamthavorn2016automated} to study more optimization approaches
in addition to the grid search. The authors investigated whether random search,
genetic algorithm, and differential evolution optimization techniques produce
significantly different results than grid search when optimizing defect
prediction models. They observed that all the investigated approaches yield similar AUC improvements after optimizing the parameters of defect
models. 
 
F. Khan et al$.$~\cite{khan2020hyper} conducted a study to assess the influence of employing an Artificial Immune Network (AIN) for hyperparameter tuning in SDP models. The investigation encompassed the exploration of eight distinct ML algorithms, such as SVM, KNN, and Random Forest. Across a selection of five datasets, the findings revealed that the utilization of AIN for hyperparameter tuning consistently resulted in performance enhancements. Specifically, it led to an increase in prediction accuracy of up to 5\% across all the studied datasets. 

Nevendra et al$.$~\cite{nevendra2022empirical} found that hyperparameter optimization of regression-based learning techniques improves the prediction performance by up to 16.96\%. The results of this study highlighted the importance of exploring the parameter space when using parameter-sensitive regression techniques.

In contrast to the mentioned prior studies that exclusively investigated the impact of hyperparameter tuning on  IVDP, our research extends this analysis to encompass a comparison with the main alternative in  SDP scenarios, the CVDP. Furthermore, our study employs a comprehensive experimental setup, incorporating an extensive array of ML algorithms and datasets commonly utilized in IVDP studies, thereby ensuring a robust evaluation of performance impacts.

\subsubsection{Cross Version Defect Prediction (CVDP)} \label{subsec:CVDPRelated}
In this SDP scenario, the training dataset for predicting defects in the current project version is trained on data extracted from the prior versions of the software~\cite{xu2019tstss}. 

Sarro et al$.$~\cite{sarro2012further} investigated the use of a Genetic Algorithm (GA) to search for a suitable configuration of SVMs to be used for CVDP. On a set of 10 software systems, Sarro et al$.$ employed a hold-out validation and used older project versions for training the models, while the most current version was used for testing. The training set used in the evaluation setting of this study contains at least two project releases. Hyperparameter tuning methods, such as Random and Grid methods, were compared with the SDP performance impact resulting from applying the  GA  algorithm. The results of this study highlighted that GA is able to effectively set SVM parameters in order to improve defect predictions within the CVDP evaluation setting. 

Fu et al.~\cite{fu2016tuning} applied optimization to the parameters of defect
prediction models based on the differential evolution (DE)
algorithm~\cite{omran2005differential}. Fu et
al.~\cite{tantithamthavorn2018impact,tantithamthavorn2016automated} used
precision and F-measure as the objectives of the DE maximization function. The
authors studied four ML algorithms and $17$ datasets from the PROMISE
repository~\cite{bowes2015different}. The results reveal that hyperparameter tuning
often leads to performance improvements and rarely worsens the performance.  Fu
et al.~\cite{fu2016tuning} used an evaluation setup that tested the SDP models on data that was not at all used in the training phase. For
example, Fu et al.~\cite{fu2016tuning} studied datasets with at least three
consecutive releases, where release $i + 1$ was built after release $i$ in
order to test their models. As such, the authors tested the built models on a
release that was not used to train the SDP models.

Nazgol et al$.$~\cite{nikravesh2023parameter} proposed a set of parameter tuners, called DEPTs, based on different variants of Differential Evolution for SDP with the Swift-Finalize strategy. To overcome issues with the conventional methods, such as the basic Differential Evolution and Grid Search, the DEPTs framework is proposed and evaluated across ten open-source projects. Nazgol et al$.$ used two project releases for training and validation, and a third release for testing. The results are highlighted using eight distinct performance metrics and found that three tuners out of five DEPTs improved prediction accuracy.

Unlike the discussed prior studies, this study focuses on comparing the performance impact caused by hyperparameter tuning across the CVDP and IVDP scenarios. This study does not focus on improving the SDP performance to higher benchmarks within a specific scenario. Hence, we highlight the possible differences between the two commonly studied SDP scenarios in the related work. Additionally, this study evaluates SDP performance on the CVDP scenario on a large scale, including 28 ML algorithms and 53 software releases. These differences should increase the generalizability and robustness of the results.

\section{Experimental Design} \label{sec:design}
 The experimental setting choices in this study are based on their popularity in prior research and applicability for SE practitioners~\cite{rajbahadur2017impact,kondo2019impact,d2010extensive,kim2011dealing,ghotra2015revisiting,zimmermann2007predicting,tantithamthavorn2016automated,tantithamthavorn2018impact,fu2016tuning, nevendra2022empirical, lee2022holistic, agrawal2021simpler}. In a comparison study like this, we selected the most common setups for IVDP and CVDP, which have been widely applied in SDP research. Additionally, we selected the frequently studied SE datasets along with underrepresented groups such as Apache Foundation projects. The included ML Algorithm families in the study cover a wide spectrum of almost all the common algorithms explored by prior research. The selected tuning algorithms have frequently been applied in SDP research as a base tuning. We believe that the setting choices of this study should improve the generalizability and impact of the findings of the results.

 In this section, we describe each setting in detail, including how we obtain the studied SE datasets, which ML algorithms we
leverage and how we evaluate the performance of the trained models across the studied SDP scenarios.  For the interested researchers, the present study can be replicated using the replication package available online\footnote{Replication package:~\url{https://github.com/DrRakha/SDP-Scenarios}.}.

\begin{table*}[!t]
	\centering
	\caption{Overview of the studied software datasets.}
		\begin{threeparttable}
			
						\begin{adjustbox}{scale=0.9}
\begin{tabular}{|c|c|c|lll|lll|}
\hline 
\multirow{2}{*}{\textsf{\textbf{\footnotesize{}Domain}}} & \multirow{2}{*}{\textsf{\textbf{\footnotesize{} Name}}} & \multirow{2}{*}{\textsf{\textbf{\footnotesize{}\#Features}}} & \multicolumn{3}{c|}{\textsf{\textbf{\footnotesize{} IVDP Dataset{*}, CVDP Training  }}} & \multicolumn{3}{c|}{\textsf{\textbf{\footnotesize{}  CVDP Testing}}}\tabularnewline
\cline{4-9} \cline{5-9} \cline{6-9} \cline{7-9} \cline{8-9} \cline{9-9} 
 &  &  & \textsf{\textbf{\footnotesize{}Pre-Release (s)}} & \textsf{\textbf{\footnotesize{}Total \#Rows}} & \textsf{\textbf{\footnotesize{}Defective (\%)}} & \textsf{\textbf{\footnotesize{}Post-Release}} & \textsf{\textbf{\footnotesize{}\#Rows}} & \textsf{\textbf{\footnotesize{}Defective (\%)}}\tabularnewline
\hline 
\multirow{6}{*}{\textsf{\footnotesize{}NASA~\cite{shepperd2013data}}} & \textsf{\scriptsize{}MC} & \textsf{\scriptsize{}39} & \textsf{\scriptsize{}1{*}} & \textsf{\scriptsize{}1,988} & \textsf{\scriptsize{}2.31\%} & \textsf{\scriptsize{}2} & \textsf{\scriptsize{}125} & \textsf{\scriptsize{}35.2\%}\tabularnewline
 & \textsf{\scriptsize{}ar} & \textsf{\scriptsize{}29} & \textsf{\scriptsize{}1{*}} & \textsf{\scriptsize{}121} & \textsf{\scriptsize{}7.44\%} & \textsf{\scriptsize{}3} & \textsf{\scriptsize{}63} & \textsf{\scriptsize{}12.6\%}\tabularnewline
 & \textsf{\scriptsize{}ar} & \textsf{\scriptsize{}29} & \textsf{\scriptsize{}1, 3{*}} & \textsf{\scriptsize{}184} & \textsf{\scriptsize{}9.24\%} & \textsf{\scriptsize{}4} & \textsf{\scriptsize{}107} & \textsf{\scriptsize{}18.7\%}\tabularnewline
 & \textsf{\scriptsize{}ar} & \textsf{\scriptsize{}29} & \textsf{\scriptsize{}1, 3, 4{*}} & \textsf{\scriptsize{}291} & \textsf{\scriptsize{}12.71\%} & \textsf{\scriptsize{}5} & \textsf{\scriptsize{}36} & \textsf{\scriptsize{}22.2\%}\tabularnewline
 & \textsf{\scriptsize{}ar} & \textsf{\scriptsize{}29} & \textsf{\scriptsize{}1, 3, 4, 5{*}} & \textsf{\scriptsize{}327} & \textsf{\scriptsize{}13.76\%} & \textsf{\scriptsize{}6} & \textsf{\scriptsize{}101} & \textsf{\scriptsize{}14.9\%}\tabularnewline
 & \textsf{\scriptsize{}PC} & \textsf{\scriptsize{}36} & \textsf{\scriptsize{} 2{*}} & \textsf{\scriptsize{}745} & \textsf{\scriptsize{}2.14\%} & \textsf{\scriptsize{}3} & \textsf{\scriptsize{}1,077} & \textsf{\scriptsize{}12.4\%}\tabularnewline
  & \textsf{\scriptsize{}PC} & \textsf{\scriptsize{}36} & \textsf{\scriptsize{} 2, 3{*}} & \textsf{\scriptsize{}1,822} & \textsf{\scriptsize{}8.23\%} & \textsf{\scriptsize{}4} & \textsf{\scriptsize{}1,287} & \textsf{\scriptsize{}13.8\%}\tabularnewline
\hline 
\multirow{36}{*}{\textsf{\footnotesize{}PROMISE~\cite{he2012investigation}}} & \textsf{\scriptsize{}ant} & \textsf{\scriptsize{}20} & \textsf{\scriptsize{}1.3{*}} & \textsf{\scriptsize{}125} & \textsf{\scriptsize{}16\%} & \textsf{\scriptsize{}1.4} & \textsf{\scriptsize{}178} & \textsf{\scriptsize{}22.5\%}\tabularnewline
 & \textsf{\scriptsize{}ant} & \textsf{\scriptsize{}20} & \textsf{\scriptsize{}1.3, 1.4{*}} & \textsf{\scriptsize{}303} & \textsf{\scriptsize{}19.8\%} & \textsf{\scriptsize{}1.5} & \textsf{\scriptsize{}293} & \textsf{\scriptsize{}10.9\%}\tabularnewline
 & \textsf{\scriptsize{}ant} & \textsf{\scriptsize{}20} & \textsf{\scriptsize{}1.3, 1.4, 1.5{*}} & \textsf{\scriptsize{}596} & \textsf{\scriptsize{}15.44\%} & \textsf{\scriptsize{}1.6} & \textsf{\scriptsize{}351} & \textsf{\scriptsize{}26.2\%}\tabularnewline
 & \textsf{\scriptsize{}ant} & \textsf{\scriptsize{}20} & \textsf{\scriptsize{}1.3, 1.4, 1.5, 1.6{*}} & \textsf{\scriptsize{}947} & \textsf{\scriptsize{}19.43\%} & \textsf{\scriptsize{}1.7} & \textsf{\scriptsize{}745} & \textsf{\scriptsize{}22.2\%}\tabularnewline
 & \textsf{\scriptsize{}camel} & \textsf{\scriptsize{}20} & \textsf{\scriptsize{}1.0{*}} & \textsf{\scriptsize{}339} & \textsf{\scriptsize{}3.83\%} & \textsf{\scriptsize{}1.2} & \textsf{\scriptsize{}608} & \textsf{\scriptsize{}35.5\%}\tabularnewline
 & \textsf{\scriptsize{}camel} & \textsf{\scriptsize{}20} & \textsf{\scriptsize{}1.0, 1.2{*}} & \textsf{\scriptsize{}947} & \textsf{\scriptsize{}24.18\%} & \textsf{\scriptsize{}1.4} & \textsf{\scriptsize{}872} & \textsf{\scriptsize{}16.6\%}\tabularnewline
 & \textsf{\scriptsize{}camel} & \textsf{\scriptsize{}20} & \textsf{\scriptsize{}1.0, 1.2, 1.4{*}} & \textsf{\scriptsize{}1,819} & \textsf{\scriptsize{}20.56\%} & \textsf{\scriptsize{}1.6} & \textsf{\scriptsize{}965} & \textsf{\scriptsize{}19.5\%}\tabularnewline
 & \textsf{\scriptsize{}forrest} & \textsf{\scriptsize{}20} & \textsf{\scriptsize{}0.6, 0.7{*}} & \textsf{\scriptsize{}35} & \textsf{\scriptsize{}17.14\%} & \textsf{\scriptsize{}0.8} & \textsf{\scriptsize{}32} & \textsf{\scriptsize{}6.3\%}\tabularnewline
 & \textsf{\scriptsize{}ivy} & \textsf{\scriptsize{}20} & \textsf{\scriptsize{}1.1{*}} & \textsf{\scriptsize{}111} & \textsf{\scriptsize{}56.76\%} & \textsf{\scriptsize{}1.4} & \textsf{\scriptsize{}241} & \textsf{\scriptsize{}6.6\%}\tabularnewline
 & \textsf{\scriptsize{}ivy} & \textsf{\scriptsize{}20} & \textsf{\scriptsize{}1.1, 1.4{*}} & \textsf{\scriptsize{}352} & \textsf{\scriptsize{}22.44\%} & \textsf{\scriptsize{}2.0} & \textsf{\scriptsize{}352} & \textsf{\scriptsize{}11.4\%}\tabularnewline
 & \textsf{\scriptsize{}jedit} & \textsf{\scriptsize{}20} & \textsf{\scriptsize{}3.2{*}} & \textsf{\scriptsize{}272} & \textsf{\scriptsize{}33.09\%} & \textsf{\scriptsize{}4.0} & \textsf{\scriptsize{}306} & \textsf{\scriptsize{}24.5\%}\tabularnewline
 & \textsf{\scriptsize{}jedit} & \textsf{\scriptsize{}20} & \textsf{\scriptsize{}3.2, 4.0{*}} & \textsf{\scriptsize{}578} & \textsf{\scriptsize{}28.55\%} & \textsf{\scriptsize{}4.1} & \textsf{\scriptsize{}312} & \textsf{\scriptsize{}25.3\%}\tabularnewline
 & \textsf{\scriptsize{}jedit} & \textsf{\scriptsize{}20} & \textsf{\scriptsize{}3.2, 4.0, 4.1{*}} & \textsf{\scriptsize{}890} & \textsf{\scriptsize{}27.42\%} & \textsf{\scriptsize{}4.2} & \textsf{\scriptsize{}367} & \textsf{\scriptsize{}13.1\%}\tabularnewline
 & \textsf{\scriptsize{}jedit} & \textsf{\scriptsize{}20} & \textsf{\scriptsize{}3.2, 4.0, 4.1, 4.2{*}} & \textsf{\scriptsize{}1,257} & \textsf{\scriptsize{}23.23\%} & \textsf{\scriptsize{}4.3} & \textsf{\scriptsize{}492} & \textsf{\scriptsize{}2.2\%}\tabularnewline
 & \textsf{\scriptsize{}xalan} & \textsf{\scriptsize{}20} & \textsf{\scriptsize{}2.4{*}} & \textsf{\scriptsize{}723} & \textsf{\scriptsize{}15.21\%} & \textsf{\scriptsize{}2.5} & \textsf{\scriptsize{}803} & \textsf{\scriptsize{}48.2\%}\tabularnewline
 & \textsf{\scriptsize{}xalan} & \textsf{\scriptsize{}20} & \textsf{\scriptsize{}2.4, 2.5{*}} & \textsf{\scriptsize{}1,526} & \textsf{\scriptsize{}32.57\%} & \textsf{\scriptsize{}2.6} & \textsf{\scriptsize{}885} & \textsf{\scriptsize{}46.4\%}\tabularnewline
 & \textsf{\scriptsize{}xalan} & \textsf{\scriptsize{}20} & \textsf{\scriptsize{}2.4, 2.5, 2.6{*}} & \textsf{\scriptsize{}2,411} & \textsf{\scriptsize{}37.66\%} & \textsf{\scriptsize{}2.7} & \textsf{\scriptsize{}909} & \textsf{\scriptsize{}98.79\%}\tabularnewline
 & \textsf{\scriptsize{}poi} & \textsf{\scriptsize{}20} & \textsf{\scriptsize{}1.5{*}} & \textsf{\scriptsize{}237} & \textsf{\scriptsize{}59.49\%} & \textsf{\scriptsize{}2.0} & \textsf{\scriptsize{}314} & \textsf{\scriptsize{}11.8\%}\tabularnewline
 & \textsf{\scriptsize{}poi} & \textsf{\scriptsize{}20} & \textsf{\scriptsize{}1.5, 2.0{*}} & \textsf{\scriptsize{}551} & \textsf{\scriptsize{}32.3\%} & \textsf{\scriptsize{}2.5} & \textsf{\scriptsize{}385} & \textsf{\scriptsize{}64.4\%}\tabularnewline
 & \textsf{\scriptsize{}poi} & \textsf{\scriptsize{}20} & \textsf{\scriptsize{}1.5, 2.0, 2.5{*}} & \textsf{\scriptsize{}936} & \textsf{\scriptsize{}45.51\%} & \textsf{\scriptsize{}3.0} & \textsf{\scriptsize{}442} & \textsf{\scriptsize{}63.6\%}\tabularnewline
 & \textsf{\scriptsize{}prop} & \textsf{\scriptsize{}20} & \textsf{\scriptsize{}1{*}} & \textsf{\scriptsize{}18,471} & \textsf{\scriptsize{}14.82\%} & \textsf{\scriptsize{}2} & \textsf{\scriptsize{}23,014} & \textsf{\scriptsize{}10.6\%}\tabularnewline
 & \textsf{\scriptsize{}prop} & \textsf{\scriptsize{}20} & \textsf{\scriptsize{}1, 2{*}} & \textsf{\scriptsize{}41,485} & \textsf{\scriptsize{}12.46\%} & \textsf{\scriptsize{}3} & \textsf{\scriptsize{}10,274} & \textsf{\scriptsize{}11.5\%}\tabularnewline
 & \textsf{\scriptsize{}prop} & \textsf{\scriptsize{}20} & \textsf{\scriptsize{}1, 2, 3{*}} & \textsf{\scriptsize{}51,759} & \textsf{\scriptsize{}12.27\%} & \textsf{\scriptsize{}4} & \textsf{\scriptsize{}8,718} & \textsf{\scriptsize{}9.6\%}\tabularnewline
 & \textsf{\scriptsize{}prop} & \textsf{\scriptsize{}20} & \textsf{\scriptsize{}1, 2, 3, 4{*}} & \textsf{\scriptsize{}60,477} & \textsf{\scriptsize{}11.89\%} & \textsf{\scriptsize{}5} & \textsf{\scriptsize{}8,516} & \textsf{\scriptsize{}15.3\%}\tabularnewline
 & \textsf{\scriptsize{}prop} & \textsf{\scriptsize{}20} & \textsf{\scriptsize{}1, 2, 3, 4, 5{*}} & \textsf{\scriptsize{}68,993} & \textsf{\scriptsize{}12.3\%} & \textsf{\scriptsize{}6} & \textsf{\scriptsize{}660} & \textsf{\scriptsize{}9.1\%}\tabularnewline
 & \textsf{\scriptsize{}log4j} & \textsf{\scriptsize{}20} & \textsf{\scriptsize{}1.0{*}} & \textsf{\scriptsize{}135} & \textsf{\scriptsize{}25.19\%} & \textsf{\scriptsize{}1.1} & \textsf{\scriptsize{}109} & \textsf{\scriptsize{}33.9\%}\tabularnewline
 & \textsf{\scriptsize{}log4j} & \textsf{\scriptsize{}20} & \textsf{\scriptsize{}1.0, 1.1{*}} & \textsf{\scriptsize{}244} & \textsf{\scriptsize{}29.1\%} & \textsf{\scriptsize{}1.2} & \textsf{\scriptsize{}189} & \textsf{\scriptsize{}92.2\%}\tabularnewline
 & \textsf{\scriptsize{}lucene} & \textsf{\scriptsize{}20} & \textsf{\scriptsize{}2.0{*}} & \textsf{\scriptsize{}195} & \textsf{\scriptsize{}46.67\%} & \textsf{\scriptsize{}2.2} & \textsf{\scriptsize{}247} & \textsf{\scriptsize{}58.3\%}\tabularnewline
 & \textsf{\scriptsize{}lucene} & \textsf{\scriptsize{}20} & \textsf{\scriptsize{}2.0, 2.2{*}} & \textsf{\scriptsize{}442} & \textsf{\scriptsize{}53.17\%} & \textsf{\scriptsize{}2.4} & \textsf{\scriptsize{}340} & \textsf{\scriptsize{}59.7\%}\tabularnewline
 & \textsf{\scriptsize{}synapse} & \textsf{\scriptsize{}20} & \textsf{\scriptsize{}1.0{*}} & \textsf{\scriptsize{}157} & \textsf{\scriptsize{}10.19\%} & \textsf{\scriptsize{}1.1} & \textsf{\scriptsize{}222} & \textsf{\scriptsize{}27\%}\tabularnewline
 & \textsf{\scriptsize{}synapse} & \textsf{\scriptsize{}20} & \textsf{\scriptsize{}1.0, 1.1{*}} & \textsf{\scriptsize{}379} & \textsf{\scriptsize{}20.05\%} & \textsf{\scriptsize{}1.2} & \textsf{\scriptsize{}256} & \textsf{\scriptsize{}33.6\%}\tabularnewline
 & \textsf{\scriptsize{}pbeans} & \textsf{\scriptsize{}20} & \textsf{\scriptsize{}1{*}} & \textsf{\scriptsize{}26} & \textsf{\scriptsize{}76.92\%} & \textsf{\scriptsize{}2} & \textsf{\scriptsize{}51} & \textsf{\scriptsize{}19.6\%}\tabularnewline
 & \textsf{\scriptsize{}velocity} & \textsf{\scriptsize{}20} & \textsf{\scriptsize{}1.4{*}} & \textsf{\scriptsize{}196} & \textsf{\scriptsize{}75\%} & \textsf{\scriptsize{}1.5} & \textsf{\scriptsize{}214} & \textsf{\scriptsize{}66.4\%}\tabularnewline
 & \textsf{\scriptsize{}velocity} & \textsf{\scriptsize{}20} & \textsf{\scriptsize{}1.4, 1.5{*}} & \textsf{\scriptsize{}410} & \textsf{\scriptsize{}70.49\%} & \textsf{\scriptsize{}1.6} & \textsf{\scriptsize{}229} & \textsf{\scriptsize{}34.1\%}\tabularnewline
 & \textsf{\scriptsize{}xerces} & \textsf{\scriptsize{}20} & \textsf{\scriptsize{}1.2{*}} & \textsf{\scriptsize{}440} & \textsf{\scriptsize{}16.14\%} & \textsf{\scriptsize{}1.3} & \textsf{\scriptsize{}453} & \textsf{\scriptsize{}15.2\%}\tabularnewline
 & \textsf{\scriptsize{}xerces} & \textsf{\scriptsize{}20} & \textsf{\scriptsize{}1.2, 1.3{*}} & \textsf{\scriptsize{}893} & \textsf{\scriptsize{}15.68\%} & \textsf{\scriptsize{}1.4} & \textsf{\scriptsize{}588} & \textsf{\scriptsize{}74.3\%}\tabularnewline
\hline 
\multirow{2}{*}{\textsf{\footnotesize{}Zimmermann~\cite{zimmermann2007predicting}}} & \textsf{\scriptsize{}eclipse} & \textsf{\scriptsize{}32} & \textsf{\scriptsize{}2.0{*}} & \textsf{\scriptsize{}6,729} & \textsf{\scriptsize{}14.49\%} & \textsf{\scriptsize{}2.1} & \textsf{\scriptsize{}7,888} & \textsf{\scriptsize{}10.8\%}\tabularnewline
 & \textsf{\scriptsize{}eclipse} & \textsf{\scriptsize{}32} & \textsf{\scriptsize{}2.0, 2.1{*}} & \textsf{\scriptsize{}14,617} & \textsf{\scriptsize{}12.51\%} & \textsf{\scriptsize{}3.0} & \textsf{\scriptsize{}10,593} & \textsf{\scriptsize{}14.8\%}\tabularnewline
\hline 
\multirow{2}{*}{\textsf{\footnotesize{}Apache~\cite{falessi2023enhancing}}} & \textsf{\scriptsize{}artemis} & \textsf{\scriptsize{}15} & \textsf{\scriptsize{}1.0{*}} & \textsf{\scriptsize{}22,708} & \textsf{\scriptsize{}3.95\%} & \textsf{\scriptsize{}2} & \textsf{\scriptsize{}30,301} & \textsf{\scriptsize{}3.03\%}\tabularnewline
 & \textsf{\scriptsize{}mng} & \textsf{\scriptsize{}15} & \textsf{\scriptsize{}1{*}} & \textsf{\scriptsize{}1,858} & \textsf{\scriptsize{}7.96\%} & \textsf{\scriptsize{}2} & \textsf{\scriptsize{}2,304} & \textsf{\scriptsize{}4.81\%}\tabularnewline
 & \textsf{\scriptsize{}mng} & \textsf{\scriptsize{}15} & \textsf{\scriptsize{}1, 2{*}} & \textsf{\scriptsize{}4,162} & \textsf{\scriptsize{}6.22\%} & \textsf{\scriptsize{}3} & \textsf{\scriptsize{}2,853} & \textsf{\scriptsize{}5.08\%}\tabularnewline
 & \textsf{\scriptsize{}mng} & \textsf{\scriptsize{}15} & \textsf{\scriptsize{}1, 2, 3{*}} & \textsf{\scriptsize{}7,015} & \textsf{\scriptsize{}5.75\%} & \textsf{\scriptsize{}4} & \textsf{\scriptsize{}3,667} & \textsf{\scriptsize{}4.28\%}\tabularnewline
  & \textsf{\scriptsize{}openjpa} & \textsf{\scriptsize{}15} & \textsf{\scriptsize{}1{*}} & \textsf{\scriptsize{}14,499} & \textsf{\scriptsize{}9.82\%} & \textsf{\scriptsize{}2} & \textsf{\scriptsize{}14,873} & \textsf{\scriptsize{}10.78\%}\tabularnewline
& \textsf{\scriptsize{}qpid} & \textsf{\scriptsize{}15} & \textsf{\scriptsize{}2{*}} & \textsf{\scriptsize{}8,800} & \textsf{\scriptsize{}4.95\%} & \textsf{\scriptsize{}3} & \textsf{\scriptsize{}9,144} & \textsf{\scriptsize{}6.46\%}\tabularnewline
& \textsf{\scriptsize{}qpid} & \textsf{\scriptsize{}15} & \textsf{\scriptsize{}2, 3{*}} & \textsf{\scriptsize{}17,944} & \textsf{\scriptsize{}5.72\%} & \textsf{\scriptsize{}4} & \textsf{\scriptsize{}9,144} & \textsf{\scriptsize{}6.02\%}\tabularnewline
& \textsf{\scriptsize{}qpid} & \textsf{\scriptsize{}15} & \textsf{\scriptsize{}2, 3, 4{*}} & \textsf{\scriptsize{}28,181} & \textsf{\scriptsize{}5.83\%} & \textsf{\scriptsize{}5} & \textsf{\scriptsize{}11,081} & \textsf{\scriptsize{}1.02\%}\tabularnewline
\hline  
\end{tabular}
			\end{adjustbox}
			\begin{tablenotes}
 
			{\small 	\item - There are 53 post-releases, including at least one release for CVDP training.
   \item - The {*} marks the software release datasets used for the IVDP scenario evaluations.
   }
 
			\end{tablenotes}
		\end{threeparttable}
\vspace*{-2mm}
	\label{tab:TableDataset1}
\end{table*}

\subsection{Studied Software Projects}\label{sec:datasets}
One criterion for the studied datasets is to select common and frequently studied datasets by SDP prior. In this study, we include datasets from publicly available software repositories that have been
thoroughly used in previous SE research for defect prediction and hyperparameter tuning~\cite{rajbahadur2017impact,kondo2019impact,d2010extensive,kim2011dealing,ghotra2015revisiting,zimmermann2007predicting,tantithamthavorn2016automated,tantithamthavorn2018impact,fu2016tuning, nevendra2022empirical, lee2022holistic, agrawal2021simpler}. Covering popular datasets allows us to directly relate the study's comparative analysis with previous SDP studies, especially on the impact of hyperparameter tuning. Also, we include datasets from underrepresented SE repositories such as Apache Foundation~\cite{falessi2023enhancing}. 
In this study, we worked on publically available 113 SE datasets\footnote{Public cleaned dataset repository provided by Tantithamthavorn~\cite{defectDataR2019}.}, including 70 datasets from the PROMISE Repository~\cite{he2012investigation},
18 datasets from NASA~\cite{shepperd2013data}, 5 datasets provided by Kim et
al.~\cite{kim2011dealing}, 5 datasets provided by D'Ambros et
al.~\cite{d2010extensive},  3 datasets provided by Zimmermann et
al.~\cite{zimmermann2007predicting}, and 12 datasets provided by Davide et al.~\cite{falessi2023enhancing} for defects on method level. We organized these software datasets in pre-release and post-release groups as illustrated in Table~\ref{tab:TableDataset1}. Out of all collected datasets, we filter out software projects that do not have at least two releases  (keeping 76 release datasets out of a total of 113 datasets available to download). In this filtration, all the datasets from  Kim et
al.~\cite{kim2011dealing} and D'Ambros et
al.~\cite{d2010extensive} are filtered. Additionally, we include underrepresented datasets with low EPV that were excluded from prior related studies~\cite{tantithamthavorn2016automated,tantithamthavorn2018impact}. Table~\ref{tab:TableDataset1} lists the datasets included in this study after that filtration.   %
In Table~\ref{tab:TableDataset1}, we describe
the number of features, number of rows, and defective percentages per studied software system after the pre-processing steps. 

This study includes roughly three times the number of datasets and includes underrepresented  included in
prior large studies on the impact of hyperparameter tuning on SDP  (i.e., $18$
datasets)~\cite{tantithamthavorn2016automated,tantithamthavorn2018impact,fu2016tuning}. Table~\ref{tab:TableDataset1} also shows the datasets used for each SDP scenario. In Table~\ref{tab:TableDataset1}, the fourth column represents pre-release software data for the CVDP scenario and the datasets for the IVDP scenario.  The last column represents the post-release for testing in the CVDP scenario.  The total number of software releases with at least one pre-release for SDP training is $53$. Following the prior research evaluations, the IVDP scenario is evaluated using one release dataset per software release cycle as the most recent pre-release  (Marked by $*$ in Table~\ref{tab:TableDataset1}).

To avoid multicollinearity issues in the tuned SDP models, we perform correlation analysis to remove highly correlated features. We use a Spearman ($\rho$) rank correlation test with a cut-off $|\rho|$ = 0.7 to identify highly correlated features~\cite{rakha2016studying,tantithamthavorn2018impact}. Furthermore, we remove redundant variables before building an SDP model. The datasets used in this study are available online within the~\texttt{DefectData}~\cite{defectDataR2019} \texttt{R} package, which provides readily accessible software defect data after applying correlation and redundancy analyses. 

\subsection{Applied ML Algorithms} 
In this study, we include 28  ML algorithms to build defect prediction models categorized across ten families.  Table~\ref{tab:Tablemodels1} lists these ten algorithm families
and the particular ML algorithms within these families.  The selection criteria for these ML algorithms can be summarized as follows:

\begin{itemize}[leftmargin=*,label={}]
	\item  \textbf{Relevance to Hyperparameter Tuning:}  Since this study aims to compare the impact of hyperparameter tuning techniques on the performance of SDP across different scenarios, we select only classification techniques that require at least one configuration parameter setting. The selected ML algorithms have varying types and quantities of hyperparameters. 
 
 \item  \textbf{Alignment with Prior Research:}  The applied ML algorithms have been widely studied in prior research across different SDP scenarios, covering different families of classification techniques. This includes the ML algorithms covered in prior research on the impact of hyperparameter tuning on SDP.

\item  \textbf{Availability and Accessibility:}  The ML algorithm implementation is available in many accessible ML libraries for SE researchers and practitioners to use for future research and applications (e.g., Caret and Weka).
\end{itemize}

To conceive a  replicable and
reproducible study, we use top-performing and popular ML algorithms (e.g., Random Forest and
Multilayer Perceptron) that were frequently explored in prior
research~\cite{kondo2019impact,fu2016tuning,tantithamthavorn2018impact,khan2020hyper, nevendra2022empirical}. To implement this study analysis, we use \texttt{R} scripting language~\cite{RpackageCite} and
\texttt{Caret} package~\cite{RcaretCite}.
To build the SDP models and apply the hyperparameters tuning, we use the \texttt{train()}
function of the \texttt{Caret} package. Table~\ref{tab:Tablemodels1} shows the parameters available for each ML algorithm along with their default values in \texttt{Caret} implementation.  To set the ML algorithm 
of the model that should be built, one has to specify the \texttt{method}
argument of the \texttt{train()} function. For example, the \texttt{method}
argument for a Random Forest algorithm is ``\texttt{rf}.'' Table~\ref{tab:Tablemodels1} shows the \texttt{method} argument for each ML algorithm in between parentheses. To apply the grid or random search tuning, we use the built-in implementation of the train function. We specify the type of hyperparameter tuning using the \texttt{search} argument. To apply default or tuned settings, we use the \texttt{tuneGrid} argument to set the setting. The argument \texttt{tuneGrid} takes a data frame with the possible search values for each tuning parameter of a certain ML algorithm, and the argument \texttt{tuneLength} is set to 5. 
For the ML algorithms with defined grid space in prior research~\cite{fu2016tuning, tantithamthavorn2018impact},  we leverage the same search space values and set them manually to the  \texttt{train} function on the argument \texttt{tuneGrid}. These values are also hard-coded in the replication package of this study.  For algorithms without such predefined grids, we rely on the \texttt{Caret} package’s automatic search space generation using the \texttt{tuneLength} parameter, which we set to $5$. In this case, \texttt{Caret} internally determines the range and granularity of candidate values based on parameter type and model-specific heuristics. For example, it applies exponential scaling for regularization parameters, linear steps for integer-valued parameters, or binary options for boolean parameters~\cite{XcaretPkg2019}. Some hyperparameters naturally have a limited search space, such as binary flags with only \texttt{TRUE} / \texttt{FALSE} values. For the default (non-tuned) setting, no search space is defined, and \texttt{Caret} uses its built-in defaults with \texttt{tuneLength} set to $1$. To maintain consistency and manage computational cost across 28 algorithms and 53 post-release datasets, we fixed tuneLength = 5 across the board (unless a manual tuneGrid was provided from prior studies ~\cite{fu2016tuning, tantithamthavorn2018impact}). This standardization allows fair comparison while keeping the search space practically bounded.

Several research studies that employed hyperparameter tuning for software prediction include a small number of ML algorithms and software datasets~\cite{wahono2014neural, gray2012reflections, fu2016tuning, osman2017hyperparameter}. Such studies are challenged by the computational power required to study the impact of hyperparameter tuning. To overcome this challenge, in this study, we leverage a high-performance computing cluster that allows our computations to cover 28 ML algorithms ({\em see} Table~\ref{tab:Tablemodels1}) and 53 post-release datasets ({\em see} Table~\ref{tab:TableDataset1}, making our study the largest in scale to date.  

\subsection{Evaluating SDP Models}\label{subsec:edm}
In this subsection, we explain the process of evaluating the SDP models. 

\textbf{Optimization Metrics:} We measure and compare the performance of SDP models when hyperparameters are tuned by using five optimization metrics: AUC, Recall Rate, Precision, F-measure, and Accuracy~\cite{zhang2014towards,boughorbel2017optimal,lessmann2008benchmarking}. These are the common metrics explored by prior research on SDP. We calculate these measures considering the {\em Defective} class as the relevant (positive) class for the performance evaluation.

\textbf{Relative Performance Impact:} \label{sec:impactPerformett}  In this study, we highlight the relative impact of hyperparameter tuning on the SDP models across the IVDP and CVDP scenarios. We measure the impact percentage by subtracting the performance under the default settings from the performance under the tuned settings as follows: 
\begin{equation}
 I(M_{\text{tuned}}, M_{\text{default}}) = \frac{M_{\text{tuned}}-M_{\text{default}}}{(M_{\text{default}}+\epsilon)} \times 100\%,
 \label{eq:perImpact1}
\end{equation}
where $M_{(\cdot)}$ is the performance measurement for a given pair of a software dataset and an ML algorithm. Specifically, the $M_{\text{tuned}}$ is the performance measured when the tuned settings are applied, while the  $M_{\text{default}}$ is the performance measured under the default settings. $\epsilon = 10^{-6}$ is used as a regularization constant (when $M_{\text{default}}=0$).   The calculated performance impact here is a percentage value that highlights the relative impact of hyperparameter tuning on the SDP performance. A positive value indicates a performance gain, while a negative value indicates a performance drop.

\textbf{Out-of-sample Bootstrap:} In this study, we leverage the Out-of-sample bootstrap~\cite{efron1983estimating} as a validation method for evaluating the trained SDP models. The choice of bootstrap as a validation method has been proven based on prior research~\cite{tantithamthavorn2016empirical} to yield the best balance between the bias and variance of estimates in the SDP context. The bootstrap validation method has been considered as standard and frequently selected in many recent SDP studies~\cite{tantithamthavorn2018impact,kondo2019impact,rajbahadur2019impact, rajbahadur2021impact, jiarpakdee2021practitioners,rajbahadur2017impact, yatish2019mining}. Bootstrap is a statistical
technique that uses random sampling with replacement~\cite{varian2005bootstrap}
for estimating errors. Suppose that we have an ML algorithm with $three$ parameters, each parameter having $5$ possible values. By using the grid search~\cite{diaz1983method}, we train $5^3=125$ models with different parameter settings via bootstrap. For each of
these 125 models, the following steps are taken:
\begin{itemize}[leftmargin=*]
	\item Randomly draw a {\em data sample} with replacement from the {\em original dataset}. The data sample has the same size as the original dataset.
	\item Train a defect prediction model on the data sample.
	\item Use the instances (from the original dataset) that do not appear in the data sample to test the model.
\end{itemize}
We repeat the bootstrap process for 100 iterations for each of the 28 studied ML algorithms, i.e., 100
data samples with replacement are drawn to generate 100 training and testing sample sets.  As we have 100 iterations, each final trained model will have 100 bootstrap-trained models trained using a training sample set, as by prior research~\cite{tantithamthavorn2018impact, gong2021revisiting}.
For each bootstrap iteration, we compute the performance measure of the
bootstrap-sampled-model on the testing sample set. In this study, we set the argument {\em metric} of \texttt{train()} function to ROC, Recall, Precision, F-measure, and Accuracy as references to the AUC, Recall Rate, Precision, F-Measure, and Accuracy performance measures. When the hyperparameter tuning is enabled, the set metric is used as the maximization objective of the tuning process.  Separate hyperparameter tuning runs are done per each included metric. 
Across all the bootstrap iterations and hyperparameter combinations, the trained model with the maximum metric value is selected to represent the testing/validation performance for IVDP and the training performance for CVDP. To perform this study experiments, we implement the out-of-sample bootstrap technique, and then we pass the training and testing indexes to the Caret \texttt{train()} function. We use the Caret function \texttt{getTrainPerf()} to get the performance summary of the bootstrap models. Figure~\ref{fig:ApproachesType} shows the bootstrap step for the IVDP and CVDP scenarios. Despite the differences between the IVDP and CVDP scenarios, the model training step in both scenarios leverages the bootstrap iterations to select the best model's hyperparameters.  In the CVDP setup, the best hyperparameters are selected using the bootstrap validation conducted only within the training releases combined data, and these selected parameters are then evaluated on a separate future release to simulate realistic deployment conditions. In contrast, the IVDP setup selects hyperparameters using bootstrap validation within the same version, with both training and testing subsets derived from a single version of the project. This allows IVDP to evaluate hyperparameter performance in a temporally static context, while CVDP reflects a more realistic temporal evolution.

\textbf{Hyperparameter Settings:} For each ML algorithm (e.g., Random Forest and Na\"{i}ve Bayes), we apply the tuned and default settings described in Section~\ref{subsec:paramSett}. For the tuned settings, we use parameter space with a size of five parameters similar to prior
research~\cite{tantithamthavorn2016automated,tantithamthavorn2018impact}. For
instance, the C5.0 algorithm has a parameter termed as \texttt{Boosting-Iterations},
which has a parameter space of values $1$, $10$, $20$, $30$, and $40$.  We use
the \texttt{tuneGrid} parameter of the \texttt{train()} function to set the
parameter space for each ML algorithm. The \texttt{train()} function produces
combinations of all parameter spaces for each ML algorithm. For example, if a
the model type has $3$ parameters, each one having $2$ values in the search array, then the total number of parameter settings to evaluate is $3^2$, which equals  $9$ possible parameter settings. 

For the scope of this study, we limit our focus to the grid search and random search (in our case, five random samples), which are frequently used algorithms for tuning a prediction model~\cite{sarro2012further, osman2017hyperparameter, tantithamthavorn2018impact}. Additionally, we based this selection decision on prior work by Tantithamthavorn et al.~\cite{tantithamthavorn2018impact}, which compared grid search with optimization techniques, such as random search, genetic algorithms, and differential evolution. Their findings showed that these approaches produced no significant differences in AUC improvements after optimizing defect prediction models' parameters. Therefore, to reduce experimental complexity, we selected the grid and random search for this study. We utilized the computational resources available by Canada Alliance clusters~\cite{CCDB} to cover the tuning of a large set of ML models and datasets. For default settings, the \texttt{tuneGrid} parameter of the \texttt{train()} function is set to include only the single max values as explained in Section~\ref{subsec:paramSett}.

\textbf{Computational Cost:} We utilize computational resources provided by the Digital Research Alliance of Canada~\cite{CCDB} to evaluate the ML algorithms studied across all datasets, performance metrics, and parameter settings. To get the results for each single performance impact combination, we run a total of 44,520 jobs (28~MLs $\times$ 53~Datasets $\times$ 5~Metrics $\times$ 3~Parameter Settings $\times$ 2~SDP Scenarios).

 \begin{figure*}[t]
	\centering
	\vspace*{-4mm}
	   \captionsetup[subfloat]{farskip=-10pt,captionskip=-2pt}
	      \captionsetup[subfigure]{labelformat=empty}
	\subfloat[\smaller \textbf{(a) Grid Tuning}]{\includegraphics[width=0.493\linewidth,keepaspectratio]{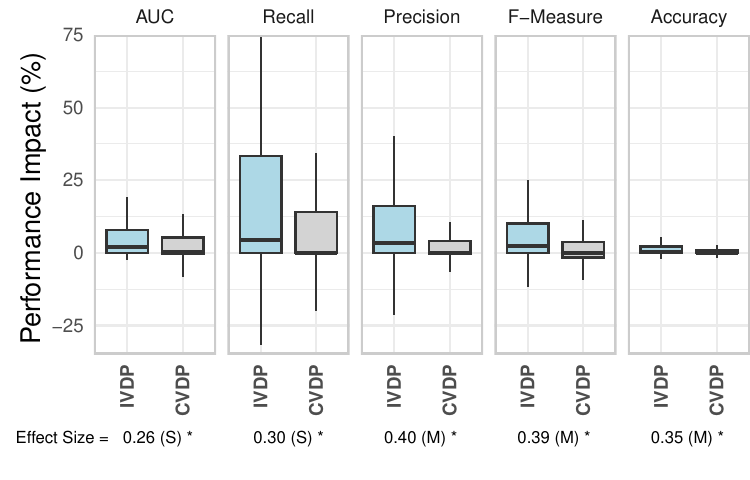}}
	\subfloat[\smaller \textbf{(b) Random Tuning}]
	{\includegraphics[width=0.493\linewidth, keepaspectratio]{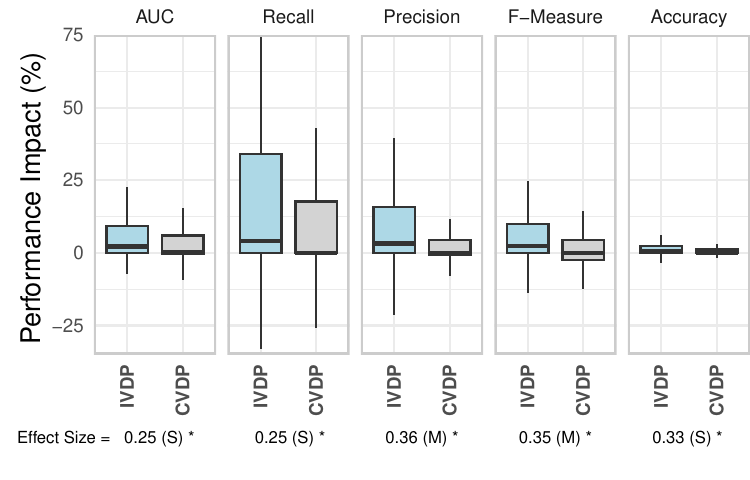}}
	
 	\captionsetup{font={normalsize}}
	\caption{\label{fig:XRQ1FigCute1} Distributions of the values of $I(M_{\text{tuned}}, M_{\text{default}})$~--- defined in the Equation~(\ref{eq:perImpact1})~--- with grid and random tunings for the IVDP and CVDP scenarios.  The $x$-axis on the plots represents the distributions' density. The $M$ in this equation for the optimized metrics starts from the AUC Impact panel on the left and ends at the Accuracy Impact panel. The black lines represent the \textbf{median} values of the distributions. The * represents statistically significant results for the Mann-Whitney U test with $p < 0.05$; $(-)$ denotes $p \ge 0.05$.  N represents \textit{negligible} effect (based on the absolute value of Cliff's delta), S~--- \textit{small} effects, M~--- \textit{medium} effect, and L~--- \textit{large} effect.  For clarity, outliers are removed from the boxplots.} 
\end{figure*}

\subsection{Improvement Perspective}\label{sec:Limitation24}

When comparing the impact of hyperparameter tuning on the performance within the IVDP and CVDP scenarios, performance improvement is a key factor~\cite{wahono2014neural, gray2012reflections,tantithamthavorn2016automated, fu2016tuning, osman2017hyperparameter, tantithamthavorn2018impact}.  %
Prior research focuses on the improvements for individual ML algorithms instead of the overall performance improvement for a defect prediction approach in a software project~\cite{wahono2014neural, gray2012reflections,tantithamthavorn2016automated, fu2016tuning, osman2017hyperparameter, tantithamthavorn2018impact}. While it is valuable to highlight the performance improvements that resulted from applying hyperparameter tuning for individual ML algorithms, specific ML algorithm improvements do not necessarily mean that hyperparameter tuning is helping the defect prediction task within the studied software project, especially if there are other well-performing ML algorithms that can be considered. For example, a performance improvement of 30\% for the C5.0-based model does not necessarily mean that the tunned C5.0 is now the best-performing ML algorithm for a software project. For instance, other ML algorithms, such as Random Forest, could have a lower performance impact but still rank as the top performer for a specific software project.  Although it is mainly a presentation perspective, it still can change the final message on the impact of hyperparameter tuning for the SDP Scenarios. In this study, we contrast the performance impact across the SDP scenarios from various perspectives, including the overall impact (i.e., RQ1), the impact per individual ML algorithms (i.e., RQ2), and dataset size (i.e., RQ3).

To statistically compare the performance impact of tuned
settings under the IVDP and CVDP scenarios, we compare the respective performance impact distributions from both scenarios. We use  {\em Mann-Whitney U} test, a non-parametric statistical test used to assess whether there is a significant difference between the two unpaired scenarios' impact~\cite{Gehan1965}. To assess the scale of the performance impact difference between the IVDP and CVDP scenarios, we use the {\em Cliff's Delta} statistic~\cite{Cliff1}. The {\em Cliff's Delta} is a non-parametric effect-size measure that indicates the magnitude of the difference between two data groups. The larger the {\em Cliff's Delta} measurement, the larger the probability that a randomly selected member from one group will be larger than a randomly selected member from the other group.  The following thresholds are used to interpret the {\em Cliff's Delta ($d$)}:
\textit{negligible~\textbf{(N)}} for $|d| \le 0.147$, \textit{small~\textbf{(S)}} for $0.147 <
|d| \le 0.33$, \textit{medium~\textbf{(M)}} for $0.33 < |d| \le 0.474$ and
\textit{large~\textbf{(L)}} for $0.474 < |d|\le 1$ (as per~\cite{Cliff2}). The statistical tests between the two scenarios' results data are considered unpaired.

 \section{Results}\label{sec:results}

\begin{figure}
	\centering
	\vspace*{-2mm}
  \includegraphics[scale=0.7]{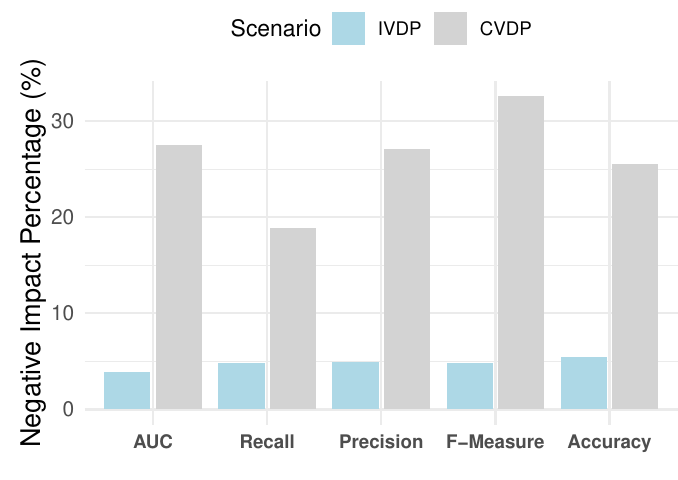} 
  \vspace*{-6mm}
	\caption{Percentage of negative impact with grid tuning for the IVDP and CVDP scenarios. The $y$-axis on the plot represents the percentage of cases where the tuning led to a negative impact on the performance across different optimization metrics.}
	\label{fig:negative}
\end{figure}

In this section, we present the motivation, approach, and results for the
research questions.
 
\subsection*{RQ1: \textbf{\RQOne}}
\noindent \textbf{Motivation.} 
The positive role of hyperparameter tuning in enhancing SDP performance is well-established~\cite{wahono2014neural,
tantithamthavorn2016automated, fu2016tuning, osman2017hyperparameter, nevendra2022empirical}, but the extent of this positive impact may vary based on the specific SDP scenario chosen for analysis and experiments. Prior research on SDP reports a diverse range of performance impacts associated with various hyperparameter tuning techniques and ML algorithms. Contrasting such impacts across distinct SDP scenarios should help assess the generalizability of reported conclusions. Moreover, it provides valuable insights for practitioners and researchers, shedding light on the expected impact of hyperparameter tuning at different stages of the software life cycle, such as transitioning from IVDP  to  CVDP. To answer this RQ, we focus on exploring how the performance impact originating from hyperparameter tuning changes across different SDP scenarios, which is not only of theoretical interest but also holds practical implications for software quality assurance.

 \begin{figure*}[t]
	\centering
	\vspace*{-4mm}
	   \captionsetup[subfloat]{farskip=-10pt,captionskip=-2pt}
	      \captionsetup[subfigure]{labelformat=empty}
	\subfloat[\smaller \textbf{(a) Negative Impact per ML Algorithm }]{\includegraphics[width=0.333\linewidth,keepaspectratio]{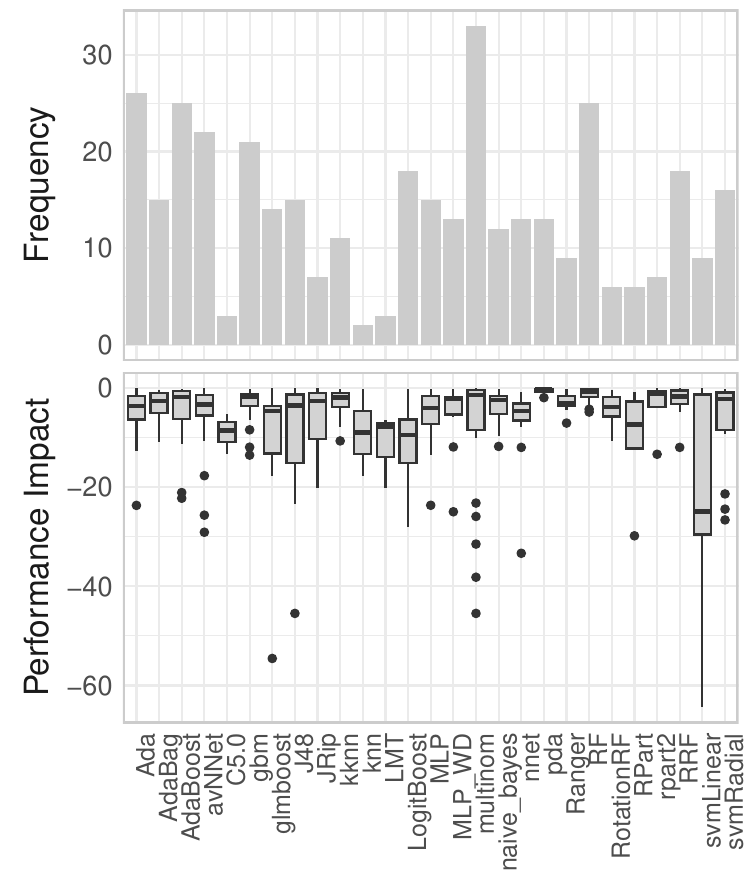}}
	\subfloat[\smaller \textbf{(b) Negative Impact per Dataset}]
	{\includegraphics[width=0.667\linewidth, keepaspectratio]{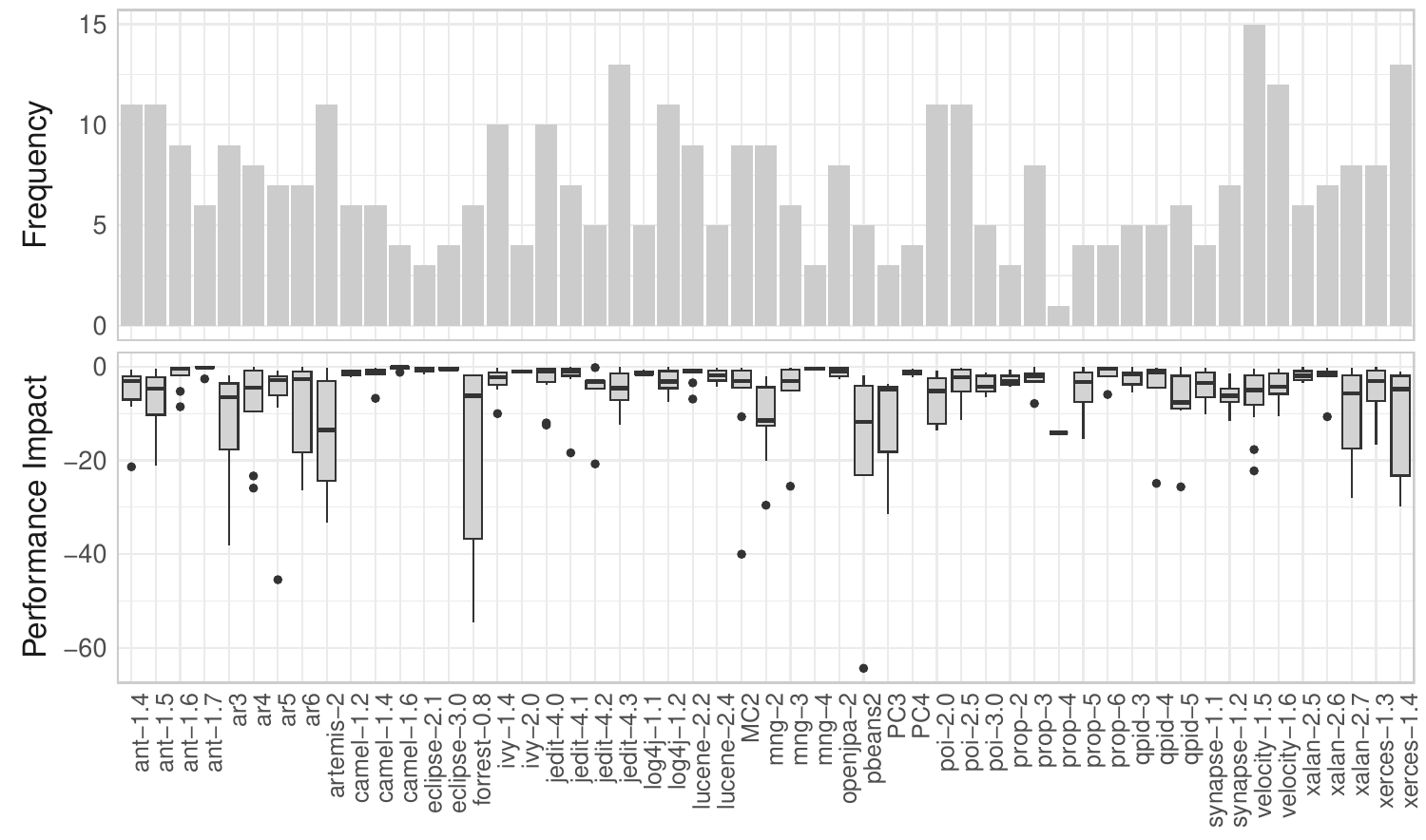}}
	
 	\captionsetup{font={normalsize}}
	\caption{\label{fig:XRQ1NegativeFigCute1} For the CVDP scenario and AUC as optimization metric, the top panels show the frequency of negative performance impacts (i.e., the number of times tuning led to a performance drop) for each ML algorithm (left) and each dataset (right). The bottom panels display the distribution of these negative impacts using boxplots. Each boxplot summarizes the magnitude and spread of negative impacts observed. This figure highlights the ML algorithms and datasets that are more prone to performance degradation when hyperparameter tuning is applied for the CVDP scenario.  } 
\end{figure*}

\noindent \textbf{Approach.}  We compare the performance impact of the 
hyperparameter tuning across the two SDP scenarios after applying the tuned and default settings. For each studied software system, we present the achieved performance across all ML algorithms and all the optimization performance metrics described in Section~\ref{subsec:edm}. To highlight the impact of hyperparameter tuning, we evaluate the impact for each performance metric using Equation~(\ref{eq:perImpact1}). For example, when the AUC metric is set as an optimization metric goal, the performance impact is referred to by \textit{AUC Impact}. The performance impact measurements for each metric are \textit{AUC Impact}, \textit{Recall Rate Impact}, \textit{Precision  Impact}, \textit{F-Measure Impact}, and \textit{Accuracy Impact}. Each pair of dataset and ML algorithm will have a single impact value, creating a distribution for each SDP scenario of $1,260$ data points (i.e., 53~post-releases $\times$ 28~ML algorithms) for an optimization performance metric. A positive value indicates that the tuned setting improves performance over the default setting by a relative percentage. A negative value indicates a performance regression when the tuned setting is applied in contrast to the default setting. We also compute the frequency percentage of negative impact for each metric and SDP scenario. To investigate how the hyperparameters change for the SDP scenario, we list the hyperparameters and their changed values after tuning.

To statistically compare the performance impact of the tuned
settings under the IVDP and CVDP evaluation setups, we compare the respective performance impact distributions. We use the {\em Mann-Whitney U} and {\em Cliff's Delta}  statistical tests described in Section~\ref{sec:Limitation24}. The null hypothesis for the Mann-Whitney Test is that there is no performance impact between the IVDP and CVDP scenarios. The alternative hypothesis is that the median of the IVDP data is greater than the median\footnote{Technically, the test also detects differences in shape and spread~\cite{hart2001mann}.} of the CVDP data. We use each ML algorithm shown in Table~\ref{tab:Tablemodels1} in the evaluations performed on each dataset.  We evaluate the IVDP and CVDP scenarios on all the software datasets shown in Table~\ref{tab:TableDataset1}. We use the \texttt{R} package {\em ggplot}  to present the performance impact distributions of each SDP scenario. We use the Equation~(\ref{eq:perImpact1}) to calculate the performance impact for each metric $M$, where we run separate hyperparameter tuning with $M$ as the optimization maximization objective.

\begin{table}

\caption{Hyperparameter Changes and Most Common Values by ML Algorithm and SDP scenario}
\centering
\begin{adjustbox}{scale=0.76}
\small
\begin{tabular}{|c|c|c|c|c|}
\hline 
\multirow{2}{*}{\textbf{Parameter - ML}} & \multicolumn{2}{c|}{\textbf{Change Freqency }-\textbf{ (\%)}} & \multicolumn{2}{c|}{\textbf{Common Value }-\textbf{ (\%)}}\tabularnewline
\cline{2-5} \cline{3-5} \cline{4-5} \cline{5-5} 
 & IVDP & CVDP & IVDP & CVDP\tabularnewline
\hline 
iter-Ada  & 13-25\% & 7-14\% & 50-75\% & 50-86\%\tabularnewline
maxdepth-Ada & 22-42.3\% & 14-28\% & 1-57.7\% & 1-72\%\tabularnewline
nu-Ada & 0-0\% & 0-0\% & 0.1-100\% & 0.1-100\%\tabularnewline
maxdepth-AdaBag & 41-82\% & 35-72.9\% & 5-64\%  & 5-56.2\%\tabularnewline
mfinal-AdaBag & 39-78\% & 37-77.1\%  & 50-22\%  & 250-37.5\%\tabularnewline
coeflearn-AdaBoost & 0-0\% & 0-0\% & Breiman-100\%  & Breiman-100\%\tabularnewline
maxdepth-AdaBoost & 35-67.3\% & 37-74\%  & 5-44.2\%  & 5-50\%\tabularnewline
mfinal-AdaBoost & 39-75\% & 37-74\%  & 250-32.7\%  & 250-32\%\tabularnewline
bag-avNNet  & 12-22.6\% & 8-17\%  & FALSE-77.4\%  & FALSE-83\%\tabularnewline
decay-avNNet & 12-79.2\% & 39-83\%  & 0.01-45.3\%  & 0.01-53.2\%\tabularnewline
size-avNNet  & 43-81.1\% & 39-83\%  & 9-41.5\%  & 9-44.7\%\tabularnewline
mode-C5.0 & 30-63.8\% & 23-47.9\% & tree-66\%  & rules-52.1\%\tabularnewline
trials-C5.0 & 25-53.2\% & 26-54.2\%  & 1-48.9\%  & 1-45.8\%\tabularnewline
winnow-C5.0 & 5-10.6\% & 10-20.8\% & FALSE-89.4\%  & FALSE-79.2\%\tabularnewline
depth-gbm & 38-76\% & 38-74.5\% & 5-50\%  & 5-56.9\%\tabularnewline
n.min-gbm  & 0-0\% & 0-0\% & 10-100\%  & 10-100\%\tabularnewline
n.trees-gbm & 47-94\% & 48-94.1\% & 250-68\%  & 250-68.6\%\tabularnewline
shrinkage-gbm & 0-0\% & 0-0\% & 0.1-100\%  & 0.1-100\%\tabularnewline
mstop-glmboost & 43-81.1\% & 44-83\% & 250-67.9\%  & 250-69.8\%\tabularnewline
prune-glmboost & 0-0\% & 0-0\% & no-100\%  & no-100\%\tabularnewline
C-J48 & 53-100\% & 50-100\% & 0.5-75.5\%  & 0.5-82\%\tabularnewline
M-J48 & 36-67.9\% & 37-74\% & 1-37.7\%  & 1-48\%\tabularnewline
MinWeights-JRip & 0-0\% & 0-0\% & 2-100\%  & 2-100\%\tabularnewline
NumFolds-JRip & 0-0\% & 0-0\% & 3-100\%  & 3-100\%\tabularnewline
Numopt-JRip & 48-90.6\% & 49-92.5\% & 5-56.6\%  & 5-71.7\%\tabularnewline
distance-kknn & 0-0\% & 0-0\% & 2-100\%  & 2-100\%\tabularnewline
kernel-kknn & 0-0\% & 0-0\% & optimal-100\%  & optimal-100\%\tabularnewline
kmax-kknn & 36-67.9\% & 37-69.8\% & 13-39.6\%  & 13-54.7\%\tabularnewline
k-knn & 12-23.1\% & 11-21.2\% & 1-76.9\%  & 1-78.8\%\tabularnewline
dimen-lda2 & 0-0\% & 0-0\% & 1-100\%  & 1-100\%\tabularnewline
iter-LMT & 46-86.8\% & 44-83\% & 81-32.1\%  & 81-41.5\%\tabularnewline
nIter-LogitBoost & 40-75.5\% & 39-73.6\% & 51-41.5\%  & 51-41.5\%\tabularnewline
size-mlp & 45-84.9\% & 44-83\% & 9-52.8\%  & 9-49.1\%\tabularnewline
decay-mlpWD & 12-22.6\% & 12-22.6\% & 0-77.4\%  & 0-77.4\%\tabularnewline
size-mlpWD & 44-83\% & 43-81.1\% & 9-54.7\%  & 9-47.2\%\tabularnewline
decay-multinom & 40-75.5\% & 44-83\% & 1e-04-26.4\%  & 1e-04-26.4\%\tabularnewline
adjust-naiveBayes & 53-100\% & 53-100\% & 1-100\%  & 1-100\%\tabularnewline
laplace-naiveBayes & 0-0\% & 0-0\% & 0-100\%  & 0-100\%\tabularnewline
kernel-naiveBayes & 7-13.2\% & 6-11.3\% & FALSE-86.8\% & FALSE-88.7\%\tabularnewline
decay-nnet & 43-81.1\% & 44-83\% & 0.001-28.3\% & 0.01-34\%\tabularnewline
size-nnet & 42-79.2\% & 42-79.2\% & 9-35.8\% & 9-47.2\%\tabularnewline
lambda-pda & 12-22.6\% & 10-18.9\% & 1-77.4\% & 1-81.1\%\tabularnewline
mtry-ranger  & 45-100\% & 36-100\% & 20-48.9\%  & 20-47.2\%\tabularnewline
size-ranger & 0-0\% & 0-0\% & 1-100\%  & 1-100\%\tabularnewline
splitrule-ranger & 16-35.6\% & 11-30.6\% & gini-64.4\%  & gini-69.4\%\tabularnewline
mtry-rf & 38-71.7\% & 36-70.6\% & 10-28.3\%  & 10-29.4\%\tabularnewline
K-rotationForest & 25-69.4\%  & 26-74.3\% & 1-30.6\%  & 2-34.3\%\tabularnewline
L-rotationForest & 15-41.7\%  & 13-37.1\% & 3-66.7\%  & 3-65.7\%\tabularnewline
cp-rpart & 43-87.8\%  & 44-89.8\% & 1e-04-32.7\%  & 1e-04-44.9\%\tabularnewline
maxdepth-rpart2 & 19-54.3\% & 21-60\% & 1-28.6\%  & 1-22.9\%\tabularnewline
coefImp-RRF & 14-36.8\%  & 20-54.1\% & 0-63.2\%  & 0-45.9\%\tabularnewline
coefReg-RRF & 34-89.5\%  & 34-91.9\% & 1-55.3\%  & 1-40.5\%\tabularnewline
mtry-RRF & 38-100\%  & 37-100\% & 2-18.4\%  & 6-24.3\%\tabularnewline
C-svmLinear & 23-47.9\%  & 22-50\%  & 1-52.1\%  & 1-50\%\tabularnewline
C-svmRadial & 51-96.2\%  & 45-97.8\% & 0.25-45.3\%  & 0.25-41.3\%\tabularnewline
sigma-svmRadial & 51-96.2\%  & 45-97.8\% & 0.1-52.8\%  & 0.1-58.7\%\tabularnewline
\hline 
\end{tabular}
\end{adjustbox}
\begin{tablenotes}
 
			{\small 	\item - These are the results with Grid Search settings.  \item - AUC is used as an optimization metric. 
            \item - The maximum change frequency per hyperparameter is 53 \\, which is the total count of tested post-release datasets}
 
			\end{tablenotes}
\label{tab:ParameterChange1}
\end{table}

\noindent \textbf{Results.} 
Figures~\ref{fig:XRQ1FigCute1} and~\ref{fig:negative}  summarize the results of this RQ and present the results of performance impact for each SDP Scenario. In Figure~\ref{fig:XRQ1FigCute1}, the $y$-axis represents the performance impact across the optimized performance metrics. The differences between the IVDP and CVDP are presented in performance impact $I$ distributions by highlighting each comparison's effect size and $p$-value.  Results of all the Mann-Whitney U tests are significant with $p<0.05$.  Figure~\ref{fig:negative} presents the frequency percentages of the negative impacts $I$ values when the hyperparameters are tuned for each SDP scenario.

\noindent \textbf{\textit{The IVDP scenario gains larger performance improvements than the CVDP scenario when hyperparameters are tuned for SDP. }}
For IVDP and CVDP scenarios, Figure~\ref{fig:XRQ1FigCute1}-a shows that the performance impacts $I$ with the grid tuning are larger for the IVDP scenario. The difference is statistically significant with non-negligible effect sizes ({\em small} 0.26 - {\em medium} 0.40) for the four optimization metrics. For example, in the grid tuning case for recall rate, the median improvement for the IVDP scenario is 4\%, whereas for CVDP, it is 0\%. Mann Whitney U test indicates that the difference is statistically significant (p $<$ 0.05); Cliff’s Delta suggests that the effect size is non-negligible ({\em small} 0.30). Despite the relatively small difference between the medians (4\% - 0\%), the deviation of values for IVDP is significantly greater than for CVDP: interquartile range (3rd quartile - 1st quartile) for IVDP is 34\% to 0.0\% while for CVDP it is 18\% to 0\%. Similar observations hold for other optimization metrics with non-negligible effect sizes varying between {\em small} and {\em medium} for all optimization metrics. The observation holds for the random search tuning with the comparison presented in Figure~\ref{fig:XRQ1FigCute1}-b, which also shows statistical differences with non-negligible effect sizes ({\em small} 0.25 - {\em medium} 0.36). These results support that the performance impact originated by hyperparameter tuning can be significantly higher just by choosing IVDP as the evaluation setup for the SDP task. It is also indicated that it is more likely for a hyperparameter tuning algorithm to increase the performance of the SDP model within the IVDP scenario over CVDP.

\noindent \textbf{\textit{The CVDP scenario gets more negative performance impacts when hyperparameters are tuned for SDP. }}
For IVDP and CVDP, Figure~\ref{fig:negative}  highlights the frequency percentage of negative impact for all the optimization performance metrics. From the results, we notice that the performance impact tends to yield more negative cases as an impact of applying hyperparameter tuning within the CVDP evaluation setup. For example, frequency percentages for the negative impact with AUC as the evaluation metric are 3.9\% and 27.4\% for IVDP and CVDP, respectively. It is interesting to see that the negative impact percentage can reach up to 32.5\% of the cases when F-Measure is selected for CVDP, in contrast to only 4.8\% for IVDP. To explore this further, Figure~\ref{fig:XRQ1NegativeFigCute1} presents the distribution and frequency of negative performance impacts specifically for the CVDP scenario across the studied ML algorithms and SE datasets. The results show that while most ML algorithms exhibit some degree of negative tuning impact on AUC, certain algorithms, such as \texttt{multinom} and \texttt{gbm}, are particularly prone to performance degradation. Likewise, some datasets consistently yield higher frequencies of negative impact. These results indicate that future studies on the impact of hyperparameter tuning should consider the negative impact of hyperparameter tuning, especially when CVDP is employed as the scenario setup. Future researchers should focus on the CVDP scenario to highlight the negative impact of a hyperparameter tuning algorithm on an SDP model.

\noindent \textbf{\textit{There are common hyperparameter values across the SDP scenarios after tuning. }}
For IVDP and CVDP, Table~\ref{tab:ParameterChange1} presents all the hyperparameters for all the studied ML algorithms, along with the frequency of changes for each SDP scenario after grid tuning, and the frequency percentage out of 53 (\%). The results show that most of the hyperparameter values get changed from default values to tuned values across both scenarios. However, some hyperparameters remain unchanged. For example, the “nu” parameters from the “ada” algorithm remain at the default value of 0.1 for all runs.

Overall, the change frequency percentage varies across the SDP scenarios, but we cannot claim which scenario has more hyperparameter variations than the other.  Table~\ref{tab:ParameterChange1}  presents the most common value per hyperparameter across the SDP scenarios. The table also shows how frequent each value is as a percentage (\%) in the Common value column. From the overall comparison, we notice that the majority of tuned hyperparameters share common values across IVDP and CVDP. However, the frequency of common values is noticeably different across the scenarios. These results indicate that the performance impact differences between IVPDP and CVDP could be associated with variations in the tuned parameters reached.

\begin{figure*}
	\centering
	\vspace*{-3mm}
  \includegraphics[scale=0.475]{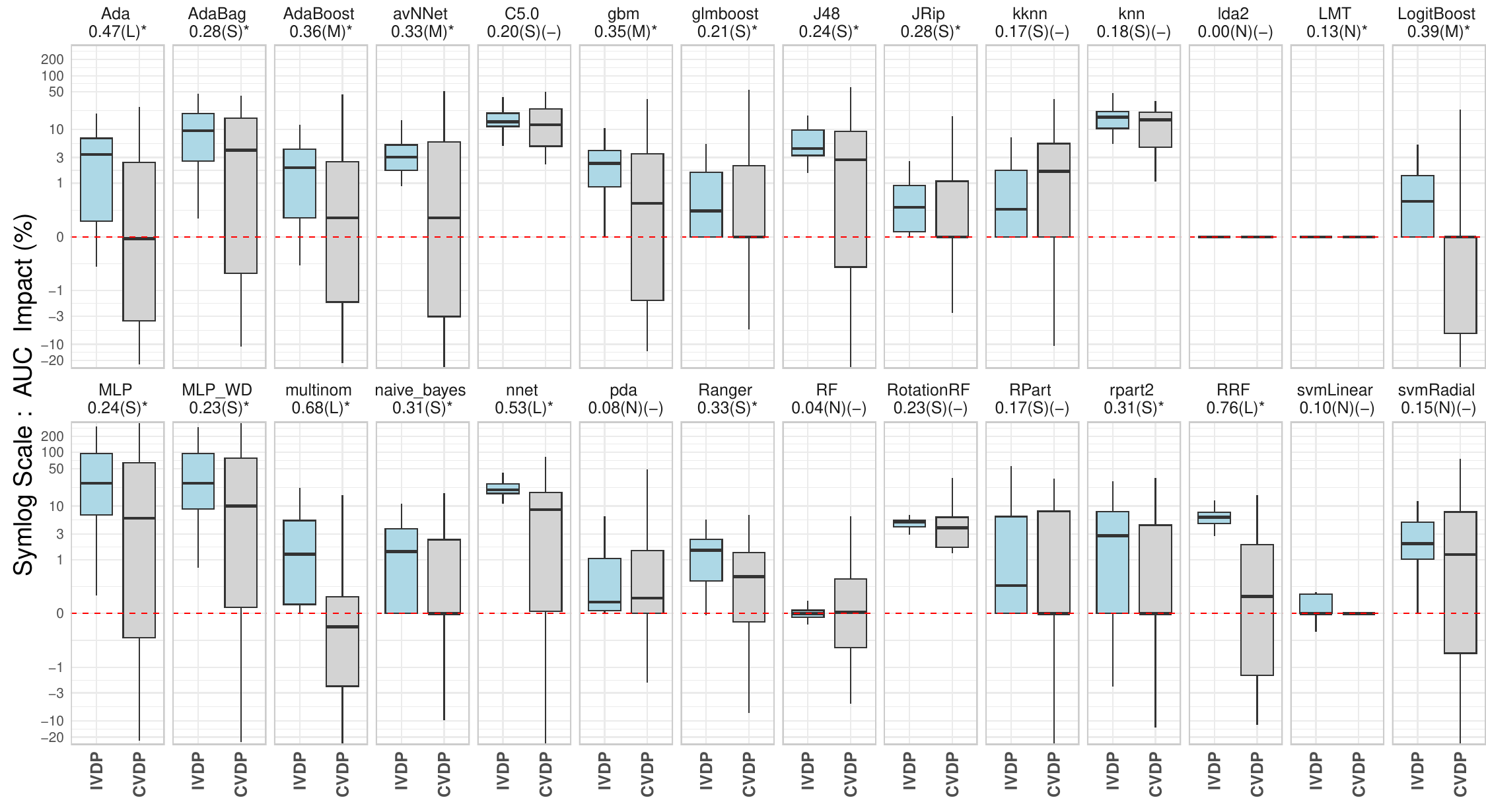} 
  \vspace*{-4mm}
	\caption{Distributions of the performance impact $I(AUC_{\text{tuned}}, AUC_{\text{default}})$ for each studied ML algorithm~--- defined in the Equation~(\ref{eq:perImpact1})~--- with grid tuning for the IVDP and CVDP Scenarios.  The $x$-axis on the plots represents the distributions' density. The black lines represent the \textbf{median} values of the distributions. The * represents statistically significant results for the Mann-Whitney U test with $p < 0.05$; $(-)$ denotes $p \ge 0.05$.  N represents \textit{negligible} effect (based on the absolute value of Cliff's delta), S~--- \textit{small} effects, M~--- \textit{medium} effect, and L~--- \textit{large} effect. The red dashed line marks the zero impact value. The y-axis uses the Symlog scale. For clarity, outliers are removed from the boxplots.}
	\label{fig:RQ2CompareAUC}
\end{figure*}

 \begin{table*}[t]
	\small
	\centering
	\caption{Scott-Knott test results when comparing the \textbf{AUC Performance Impact} $I(AUC_{\text{tuned}}, AUC_{\text{default}})$ ranking per ML algorithm for both SDP scenarios. The rankings
are divided into groups that have a statistically significant difference in the performance impact mean.}
		\vspace*{-2mm}
	\begin{adjustbox}{scale=0.905}
		\begin{threeparttable}
\begin{tabular}{|c|c|c|c|c|c|c|c|c|c|c|c|}
\hline 
\multicolumn{6}{|c|}{Ranking Based on \textbf{Grid} Hyperparameter Tuning} & \multicolumn{6}{c|}{Ranking Based on \textbf{Random} Hyperparameter Tuning}\tabularnewline
\hline 
\multicolumn{3}{|c|}{\textbf{IVDP}} & \multicolumn{3}{c|}{\textbf{CVDP}} & \multicolumn{3}{c|}{\textbf{IVDP}} & \multicolumn{3}{c|}{\textbf{CVDP}}\tabularnewline
\hline 
Group & ML Model & Mean & Group & ML Model & Mean & Group & ML Model & Mean & Group & ML Model & Mean\tabularnewline
\hline 
\multirow{2}{*}{1} & mlpWD & 60.02  & \multirow{2}{*}{1} & mlpWD & 56.88 & \multirow{2}{*}{1} & mlpWD & 60.01 & \multirow{2}{*}{1} & mlpWD & 56.89\tabularnewline
\cline{2-3} \cline{3-3} \cline{5-6} \cline{6-6} \cline{8-9} \cline{9-9} \cline{11-12} \cline{12-12} 
 & mlp & 58.39  &  & mlp & 55.59 &  & mlp & 58.40 &  & mlp & 55.60\tabularnewline
\hline 
2 & nnet  & 22.76 & \multirow{3}{*}{2} & C5.0  & 14.05 & 2 & nnet  & 22.77 & \multirow{3}{*}{2} & C5.0  & 14.05\tabularnewline
\cline{1-3} \cline{2-3} \cline{3-3} \cline{5-9} \cline{6-9} \cline{7-9} \cline{8-9} \cline{9-9} \cline{11-12} \cline{12-12} 
\multirow{2}{*}{3} & C5.0  & 17.46 &  & knn & 13.66  & \multirow{2}{*}{3} & C5.0  & 17.47 &  & knn  & 13.67\tabularnewline
\cline{2-3} \cline{3-3} \cline{5-6} \cline{6-6} \cline{8-9} \cline{9-9} \cline{11-12} \cline{12-12} 
 & knn & 16.95 &  & nnet  & 11.92  &  & knn & 16.95 &  & nnet  & 11.92\tabularnewline
\hline 
4 & AdaBag  & 12.65 & 3 & AdaBag & 7.97 & \multirow{2}{*}{4} & kknn & 13.37 & \multirow{3}{*}{3} & kknn & 8.87\tabularnewline
\cline{1-6} \cline{2-6} \cline{3-6} \cline{4-6} \cline{5-6} \cline{6-6} \cline{8-9} \cline{9-9} \cline{11-12} \cline{12-12} 
\multirow{3}{*}{5} & RPart  & 7.37 & \multirow{4}{*}{4} & J48  & 5.09 &  & AdaBag & 12.68 &  & AdaBag & 8.16\tabularnewline
\cline{2-3} \cline{3-3} \cline{5-9} \cline{6-9} \cline{7-9} \cline{8-9} \cline{9-9} \cline{11-12} \cline{12-12} 
 & RRF  & 6.88 &  & RotationRF & 4.64  & \multirow{2}{*}{5} & J48 & 8.07 &  & J48 & 7.82\tabularnewline
\cline{2-3} \cline{3-3} \cline{5-6} \cline{6-6} \cline{8-12} \cline{9-12} \cline{10-12} \cline{11-12} \cline{12-12} 
 & J48  & 5.96 &  & kknn  & 3.86 &  & RPart & 7.36 & 4 & RotationRF  & 5.77\tabularnewline
\cline{1-3} \cline{2-3} \cline{3-3} \cline{5-12} \cline{6-12} \cline{7-12} \cline{8-12} \cline{9-12} \cline{10-12} \cline{11-12} \cline{12-12} 
\multirow{3}{*}{6} & RotationRF & 4.98 &  & svmRadial  & 3.68 & 6 & RotationRF  & 6.11 & \multirow{4}{*}{5} & avNNet & 3.59\tabularnewline
\cline{2-9} \cline{3-9} \cline{4-9} \cline{5-9} \cline{6-9} \cline{7-9} \cline{8-9} \cline{9-9} \cline{11-12} \cline{12-12} 
 & rpart2 & 4.92 & \multirow{4}{*}{5} & RPart  & 3.35  & \multirow{2}{*}{7} & rpart2  & 5.26 &  & svmRadial & 3.54\tabularnewline
\cline{2-3} \cline{3-3} \cline{5-6} \cline{6-6} \cline{8-9} \cline{9-9} \cline{11-12} \cline{12-12} 
 & avNNet & 4.43 &  & avNNet  & 2.59  &  & avNNet & 4.42 &  & RPart  & 3.35\tabularnewline
\cline{1-3} \cline{2-3} \cline{3-3} \cline{5-9} \cline{6-9} \cline{7-9} \cline{8-9} \cline{9-9} \cline{11-12} \cline{12-12} 
\multirow{3}{*}{7} & Ada  & 4.19 &  & rpart2  & 2.54  & \multirow{2}{*}{8} & multinom & 3.70 &  & rpart2  & 2.83\tabularnewline
\cline{2-3} \cline{3-3} \cline{5-6} \cline{6-6} \cline{8-12} \cline{9-12} \cline{10-12} \cline{11-12} \cline{12-12} 
 & multinom & 3.69 &  & pda  & 2.04  &  & svmRadial & 3.48 & \multirow{2}{*}{6} & pda  & 2.04\tabularnewline
\cline{2-9} \cline{3-9} \cline{4-9} \cline{5-9} \cline{6-9} \cline{7-9} \cline{8-9} \cline{9-9} \cline{11-12} \cline{12-12} 
 & svmRadial & 3.56 & \multirow{8}{*}{6} & gbm  & 1.40  & \multirow{5}{*}{9} & RRF & 2.95 &  & gbm & 1.40\tabularnewline
\cline{1-3} \cline{2-3} \cline{3-3} \cline{5-6} \cline{6-6} \cline{8-12} \cline{9-12} \cline{10-12} \cline{11-12} \cline{12-12} 
\multirow{2}{*}{8} & AdaBoost  & 2.90 &  & RRF  & 1.13  &  & Ada  & 2.88 & \multirow{6}{*}{7} & Ranger & 1.20\tabularnewline
\cline{2-3} \cline{3-3} \cline{5-6} \cline{6-6} \cline{8-9} \cline{9-9} \cline{11-12} \cline{12-12} 
 & gbm & 2.78 &  & LMT  & 0.75 &  & gbm & 2.78 &  & AdaBoost  & 0.94\tabularnewline
\cline{1-3} \cline{2-3} \cline{3-3} \cline{5-6} \cline{6-6} \cline{8-9} \cline{9-9} \cline{11-12} \cline{12-12} 
9 & naive\_bayes & 2.30 &  & naive\_bayes  & 0.73  &  & AdaBoost  & 2.65 &  & RRF  & 0.90\tabularnewline
\cline{1-3} \cline{2-3} \cline{3-3} \cline{5-6} \cline{6-6} \cline{8-9} \cline{9-9} \cline{11-12} \cline{12-12} 
10 & Ranger  & 1.73 &  & Ranger  & 0.61 &  & naive\_bayes  & 2.35 &  & LMT & 0.75\tabularnewline
\cline{1-3} \cline{2-3} \cline{3-3} \cline{5-9} \cline{6-9} \cline{7-9} \cline{8-9} \cline{9-9} \cline{11-12} \cline{12-12} 
11 & kknn & 1.20 &  & AdaBoost  & 0.59  & \multirow{3}{*}{10} & Ranger  & 1.19 &  & naive\_bayes & 0.74\tabularnewline
\cline{1-3} \cline{2-3} \cline{3-3} \cline{5-6} \cline{6-6} \cline{8-9} \cline{9-9} \cline{11-12} \cline{12-12} 
\multirow{3}{*}{12} & LogitBoost  & 1.02 &  & glmboost  & 0.46  &  & glmboost & 1.05 &  & JRip & 0.43\tabularnewline
\cline{2-3} \cline{3-3} \cline{5-6} \cline{6-6} \cline{8-12} \cline{9-12} \cline{10-12} \cline{11-12} \cline{12-12} 
 & glmboost & 1.00 &  & JRip  & 0.43  &  & LogitBoost  & 1.02 & \multirow{3}{*}{8} & lda2  & 0.00\tabularnewline
\cline{2-9} \cline{3-9} \cline{4-9} \cline{5-9} \cline{6-9} \cline{7-9} \cline{8-9} \cline{9-9} \cline{11-12} \cline{12-12} 
 & pda  & 0.84 & \multirow{3}{*}{7} & Ada  & 0.02 & 11 & pda  & 0.84 &  & RF  & -0.05\tabularnewline
\cline{1-3} \cline{2-3} \cline{3-3} \cline{5-9} \cline{6-9} \cline{7-9} \cline{8-9} \cline{9-9} \cline{11-12} \cline{12-12} 
\multirow{3}{*}{13} & JRip & 0.66 &  & lda2  & 0.00 & \multirow{3}{*}{12} & JRip  & 0.66 &  & glmboost  & -0.27\tabularnewline
\cline{2-3} \cline{3-3} \cline{5-6} \cline{6-6} \cline{8-12} \cline{9-12} \cline{10-12} \cline{11-12} \cline{12-12} 
 & LMT  & 0.55 &  & RF  & -0.09 &  & svmLinear & 0.55 & \multirow{3}{*}{9} & LogitBoost  & -1.21\tabularnewline
\cline{2-6} \cline{3-6} \cline{4-6} \cline{5-6} \cline{6-6} \cline{8-9} \cline{9-9} \cline{11-12} \cline{12-12} 
 & svmLinear  & 0.53 & \multirow{2}{*}{8} & LogitBoost  & -1.21 &  & LMT  & 0.54 &  & svmLinear & -1.54\tabularnewline
\cline{1-3} \cline{2-3} \cline{3-3} \cline{5-9} \cline{6-9} \cline{7-9} \cline{8-9} \cline{9-9} \cline{11-12} \cline{12-12} 
\multirow{2}{*}{14} & lda2  & 0.00 &  & svmLinear  & -1.51 & \multirow{2}{*}{13} & lda2  & 0.00 &  & Ada & -1.63\tabularnewline
\cline{2-6} \cline{3-6} \cline{4-6} \cline{5-6} \cline{6-6} \cline{8-12} \cline{9-12} \cline{10-12} \cline{11-12} \cline{12-12} 
 & RF  & -0.01 & 9 & multinom & -3.58  &  & RF & -0.01 & 10 & multinom  & -3.58\tabularnewline
\hline 
\end{tabular}
		\end{threeparttable}
	\end{adjustbox}
	\label{tab:rq2ranks1}
 \vspace*{-4mm}
\end{table*}

 \begin{table*}[t]
	\small
	\centering
	\caption{Scott-Knott test results when comparing the actual \textbf{AUC Performance} ranking per ML algorithm across all datasets for both SDP scenarios. The table show the ranking before and after applying the Grid Tuning. The rankings
are divided into groups that have a statistically significant difference in the AUC performance mean. }
		\vspace*{-2mm}
	\begin{adjustbox}{scale=0.845}
		\begin{threeparttable}
\begin{tabular}{|c|c|c|c|c|c|c|c|c|c|c|c|}
\hline 
\multicolumn{6}{|c|}{IVDP } & \multicolumn{6}{c|}{CVDP}\tabularnewline
\hline 
\multicolumn{3}{|c|}{\textbf{Before Tunning}} & \multicolumn{3}{c|}{\textbf{After Tunning}} & \multicolumn{3}{c|}{\textbf{Before Tunning}} & \multicolumn{3}{c|}{\textbf{After Tunning}}\tabularnewline
\hline 
Group & ML Algorithm & Mean & Group & ML Algorithm & Mean & Group & ML Algorithm & Mean & Group & ML Algorithm & Mean\tabularnewline
\hline 
\multirow{5}{*}{1} & RF & 0.745 & \multirow{3}{*}{1} & AdaBoost & 0.755 & \multirow{4}{*}{1} & naive\_bayes & 0.696 & \multirow{6}{*}{1} & naive\_bayes & 0.700\tabularnewline
\cline{2-3} \cline{3-3} \cline{5-6} \cline{6-6} \cline{8-9} \cline{9-9} \cline{11-12} \cline{12-12} 
 & glmboost & 0.743 &  & Ada & 0.750 &  & AdaBoost & 0.688 &  & AdaBoost & 0.690\tabularnewline
\cline{2-3} \cline{3-3} \cline{5-6} \cline{6-6} \cline{8-9} \cline{9-9} \cline{11-12} \cline{12-12} 
 & pda & 0.739 &  & glmboost  & 0.750 &  & LMT & 0.682 &  & MLP\_WD & 0.687\tabularnewline
\cline{2-6} \cline{3-6} \cline{4-6} \cline{5-6} \cline{6-6} \cline{8-9} \cline{9-9} \cline{11-12} \cline{12-12} 
 & AdaBoost & 0.734 & \multirow{7}{*}{2} & RF & 0.745  &  & Ada & 0.681 &  & LMT & 0.686\tabularnewline
\cline{2-3} \cline{3-3} \cline{5-9} \cline{6-9} \cline{7-9} \cline{8-9} \cline{9-9} \cline{11-12} \cline{12-12} 
 & naive\_bayes & 0.725 &  & pda & 0.745 & \multirow{3}{*}{2} & RF & 0.677 &  & pda & 0.684\tabularnewline
\cline{1-3} \cline{2-3} \cline{3-3} \cline{5-6} \cline{6-6} \cline{8-9} \cline{9-9} \cline{11-12} \cline{12-12} 
\multirow{4}{*}{2} & Ada & 0.720 &  & MLP\_WD & 0.744 &  & pda & 0.673 &  & nnet & 0.683\tabularnewline
\cline{2-3} \cline{3-3} \cline{5-6} \cline{6-6} \cline{8-12} \cline{9-12} \cline{10-12} \cline{11-12} \cline{12-12} 
 & lda2 & 0.719 &  & naive\_bayes & 0.742 &  & lda2 & 0.661 & \multirow{3}{*}{2} & Ada & 0.679\tabularnewline
\cline{2-3} \cline{3-3} \cline{5-9} \cline{6-9} \cline{7-9} \cline{8-9} \cline{9-9} \cline{11-12} \cline{12-12} 
 & LMT  & 0.713 &  & nnet & 0.741 & 3 & gbm & 0.651 &  & RF & 0.676\tabularnewline
\cline{2-3} \cline{3-3} \cline{5-9} \cline{6-9} \cline{7-9} \cline{8-9} \cline{9-9} \cline{11-12} \cline{12-12} 
 & multinom & 0.712 &  & multinom & 0.737 & \multirow{4}{*}{4} & AdaBag & 0.626 &  & MLP & 0.670\tabularnewline
\cline{1-3} \cline{2-3} \cline{3-3} \cline{5-6} \cline{6-6} \cline{8-12} \cline{9-12} \cline{10-12} \cline{11-12} \cline{12-12} 
\multirow{2}{*}{3} & gbm & 0.703 &  & MLP & 0.735 &  & avNNet & 0.625 & \multirow{6}{*}{3} & AdaBag & 0.668\tabularnewline
\cline{2-6} \cline{3-6} \cline{4-6} \cline{5-6} \cline{6-6} \cline{8-9} \cline{9-9} \cline{11-12} \cline{12-12} 
 & LogitBoost & 0.693 & \multirow{5}{*}{3} & AdaBag & 0.725 &  & kknn & 0.625 &  & lda2 & 0.668\tabularnewline
\cline{1-3} \cline{2-3} \cline{3-3} \cline{5-6} \cline{6-6} \cline{8-9} \cline{9-9} \cline{11-12} \cline{12-12} 
\multirow{3}{*}{4} & avNNet & 0.669 &  & gbm & 0.722 &  & nnet & 0.617 &  & gbm & 0.658\tabularnewline
\cline{2-3} \cline{3-3} \cline{5-9} \cline{6-9} \cline{7-9} \cline{8-9} \cline{9-9} \cline{11-12} \cline{12-12} 
 & svmRadial & 0.658 &  & knn & 0.721 & \multirow{3}{*}{5} & JRip & 0.582 &  & knn & 0.657\tabularnewline
\cline{2-3} \cline{3-3} \cline{5-6} \cline{6-6} \cline{8-9} \cline{9-9} \cline{11-12} \cline{12-12} 
 & AdaBag & 0.651 &  & lda2 & 0.719 &  & svmRadial & 0.579 &  & kknn & 0.648\tabularnewline
\cline{1-3} \cline{2-3} \cline{3-3} \cline{5-6} \cline{6-6} \cline{8-9} \cline{9-9} \cline{11-12} \cline{12-12} 
5 & kknn & 0.633 &  & LMT & 0.717 &  & knn & 0.579 &  & avNNet & 0.633\tabularnewline
\hline 
\end{tabular}
		\end{threeparttable}
	\end{adjustbox}
	\label{tab:rq2ranksMLAlgorithhm}
 \vspace*{-4mm}
\end{table*}

\noindent \textbf{Discussion.} The results of this RQ statistically show that the studied SDP scenarios significantly differ regarding the performance impact resulting from applying hyperparameter tuning.   Overall, the obvious reason for the poor performance impact is that some ML Algorithms' performance does not benefit from the hyperparameter tuning like others, leading to a small performance impact and sometimes a negative impact.  This study focuses on the performance impact resulting from applying hyperparameter tuning while other factors (e.g., data splits, random seeds, setup, and model features) remain constant between the tuned settings and the default settings. The only difference is the SDP scenario that is applied for comparison. These results suggested that the performance impact reported for IVDP~\cite{wahono2014neural,osman2017hyperparameter,tantithamthavorn2018impact,agrawal2018better,ali2021empirical} scenarios studies does not necessarily generalize for the CVDP~\cite{fu2016differential,fu2016tuning, lee2022holistic, agrawal2021simpler}, indicating that the impact of tuning is scenario-dependent. 

As the IVDP selects the best-performing tuned model from the bootstrap validation, it is less likely that the tuned model will underperform the default model. Conversely, the CVDP parameter settings are selected on training releases. Testing is then applied to a new release. To obtain more robust results, we recommend examining hyperparameter tuning algorithms across multiple SDP scenarios in future SE studies. As CVDP is perceived as the practical scenario for later stages of the software life cycle~\cite{fu2016tuning,nikravesh2023parameter}, we recommend future research to investigate further the negative performance impacts of applying hyperparameter tuning techniques on SDP performance, as SE practitioners may be misled by focusing solely on the positive impact of tuning hyperparameters.

While this study does not aim to predict when tuning may have a negative impact on SDP performance, the study results provide a practical starting point for further investigation. We fully acknowledge that understanding the conditions under which hyperparameter tuning leads to negative impacts, such as differences in data distributions, sensitivity to ML algorithmic changes, or SE dataset characteristics, is a critical direction for future work. Investigating these issues will require new methodological approaches, such as analyzing the relationship between concept drift and tuning robustness, or identifying characteristics  of ML algorithms and SE datasets that are more prone to performance degradation after tuning~\cite{gangwar2023concept, kabir2019assessing}.

To support this future research direction, RQ1  analysis provides a detailed account of where negative tuning effects are spread. By examining 28 ML algorithms and 53 datasets, our study helps pinpoint specific conditions under which tuning harms performance, particularly under CVDP. We encourage future studies to build upon these findings and develop adaptive or robust tuning strategies that mitigate these effects in practical SDP deployments.

\conclusionbox{\textbf{Summary:} The results show that the performance impact of applying hyperparameter tuning significantly varies across SDP scenarios. This suggests that tuning benefits observed in one scenario (e.g., IVDP) may not generalize to others (e.g., CVDP).
Future SE researchers and practitioners should examine tuning impact across multiple scenarios to ensure practical relevance and investigate the mitigations of negative performance impacts. }

\subsection*{RQ2: \textbf{\RQTwo}}
\vspace*{-1mm}
\noindent \textbf{Motivation.} 
This RQ builds upon prior research by delving into the nuanced relationship between hyperparameter tuning and selecting the top ML algorithm in the SDP scenario. The quest for the best-performing ML algorithms becomes particularly crucial after applying hyperparameter tuning, which significantly impacts the SDP model performance. Consequently, the comparative analysis of ML algorithms rankings across two primary SDP scenarios is a cornerstone for deriving insightful and scenario-specific conclusions. Identifying the best-performing ML algorithms enhances the likelihood that the SDP system will perform well in diverse settings, improving its applicability across various software development scenarios.

\noindent \textbf{Approach.} 
In this RQ, we compare the performance impact of hyperparameter tuning from the perspective of each  ML algorithm across the studied SDP scenarios. To investigate how the performance impact differs between the scenarios across each ML algorithm, we apply the same statistical tests as RQ1, including {\em Mann-Whitney U}  test and {\em Cliff's Delta} effect-size, to contrast the performance impact defined in Equation~(\ref{eq:perImpact1}). This comparison should highlight which ML algorithm's impact differs significantly between the two SDP scenarios.  To compute a statistically stable ranking of the performance impact per the ML algorithms, we use the \textit{Scott-Knott} test~\cite{tantithamthavorn2015impact} (with $p < 0.05$) across the studied SDP scenarios. The Scott-Knott test is a statistical method for multi-comparison cluster analysis. It groups the factors based on their reported importance values across ten folds. If the difference between two clusters is statistically significant, the Scott-Knott test segregates the factors into distinct clusters.  The Scott-Knott test runs recursively until no more clusters can be created. To understand how the importance of different ML model features changes due to the hyperparameter tuning, we evaluate the variable importance difference for every feature before and after the tuning for both SDP scenarios. This means that for every ML algorithm, dataset, and SDP scenario, a feature will have a difference in importance value, highlighting the change in that analysis metric for SDP modeling.

 \begin{figure*}[t]
	\centering
       \label{fig:XVarImport}
	\vspace*{-4mm}
	   \captionsetup[subfloat]{farskip=-10pt,captionskip=-2pt}
	      \captionsetup[subfigure]{labelformat=empty}
	\subfloat[\smaller \textbf{(a) IVDP Feature Importance Difference}]{\includegraphics[width=0.495\linewidth,keepaspectratio]{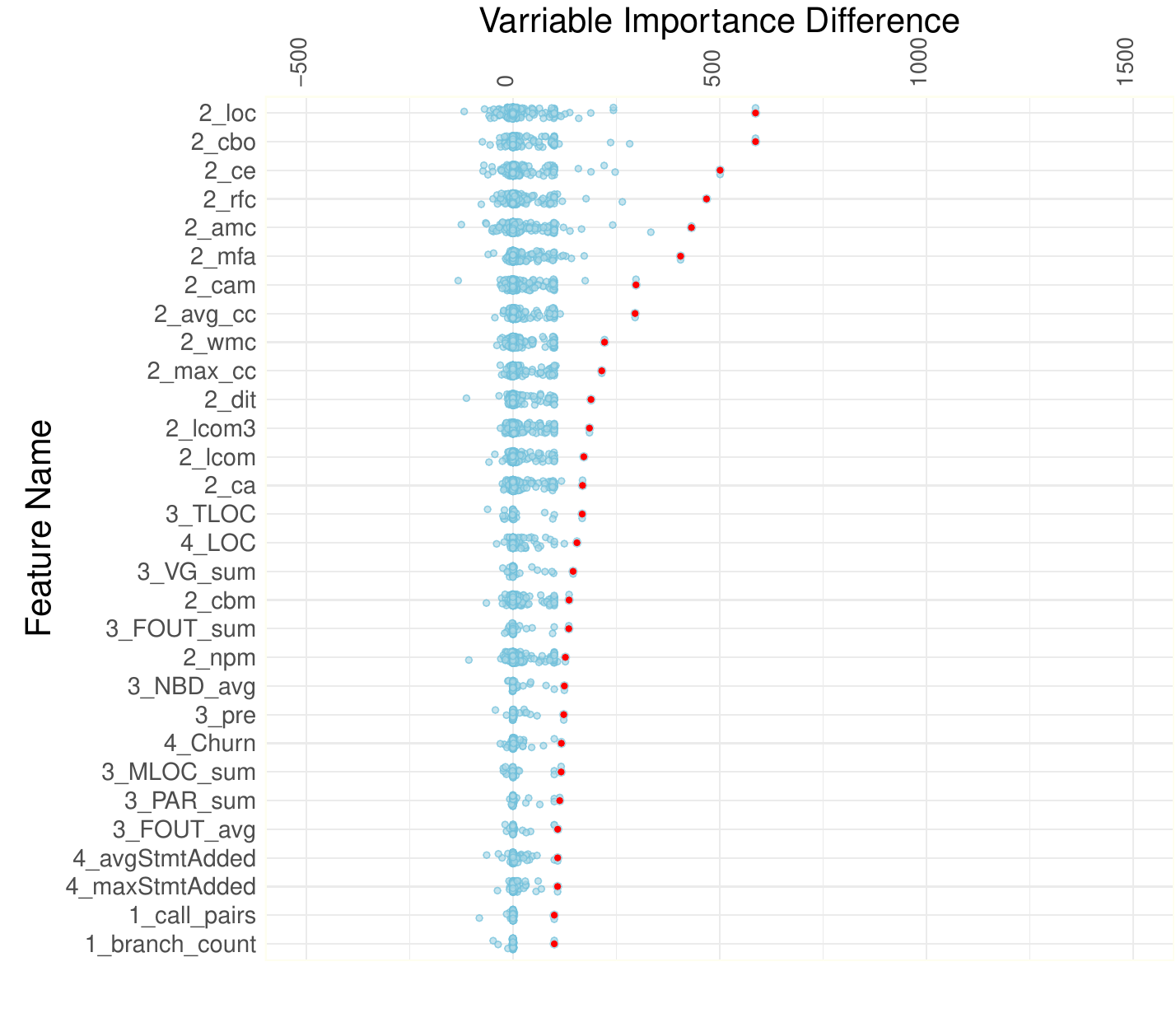}}
	\subfloat[\smaller \textbf{(b) CVDP Feature Importance Difference }]
	{\includegraphics[width=0.495\linewidth, keepaspectratio]{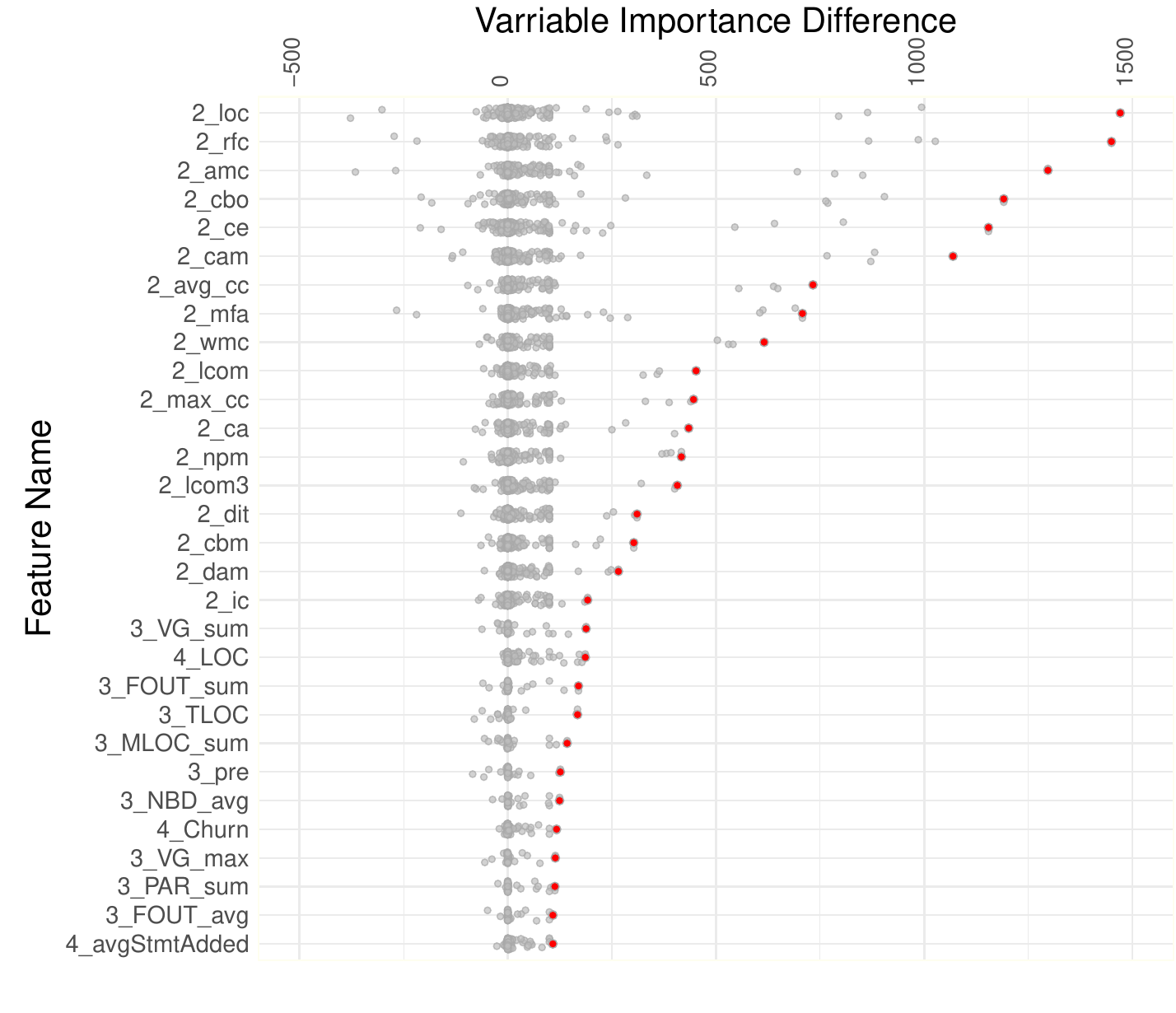}}
	
 	\captionsetup{font={normalsize}}
	\caption{ Distribution of feature importance differences for all features under the IVDP and CVDP  scenarios with grid as tuning and ROC as optimization metric. The $x$-axis represents the feature important difference values with a range limited to $-500$ to $1000$. The $y$-axis lists the top 30 features out of a total of 135 features ranked by the maximum importance difference. The prefix number refers to the dataset group ($1\_$ maps to NASA, $2\_$ --- to PROMISE, $3\_$ --- to Zimmermann, and $4\_$ --- to Apache). The figure shows the distribution of all differences within the specified range per each feature. The maximum importance difference for each feature is highlighted with a red point.} 
\end{figure*}

\begin{figure*}[h]
	\centering
	\vspace*{-2mm}
	   \captionsetup[subfloat]{farskip=-2pt,captionskip=-5pt}
	      \captionsetup[subfigure]{labelformat=empty}
	\subfloat[\smaller \textbf{(a) Small Training ( $\leq$ Median  ) }]{\includegraphics[width=0.480\linewidth,keepaspectratio]{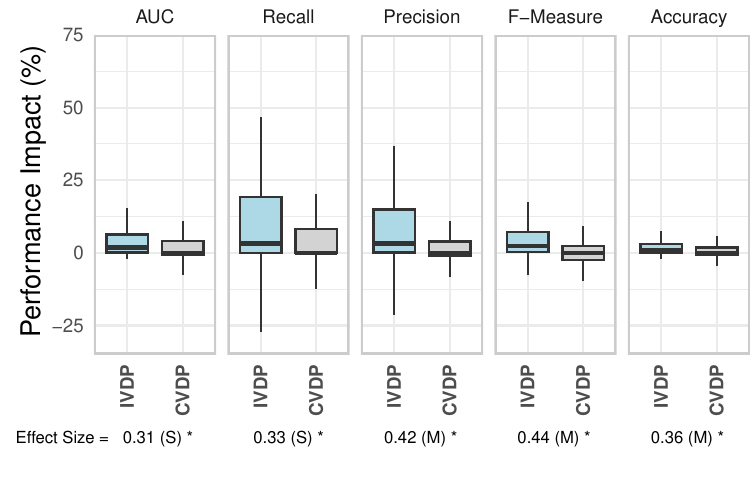}}
	\subfloat[\smaller \textbf{(b) Large Training ( $>$ Median)}]
	{\includegraphics[width=0.480\linewidth, keepaspectratio]{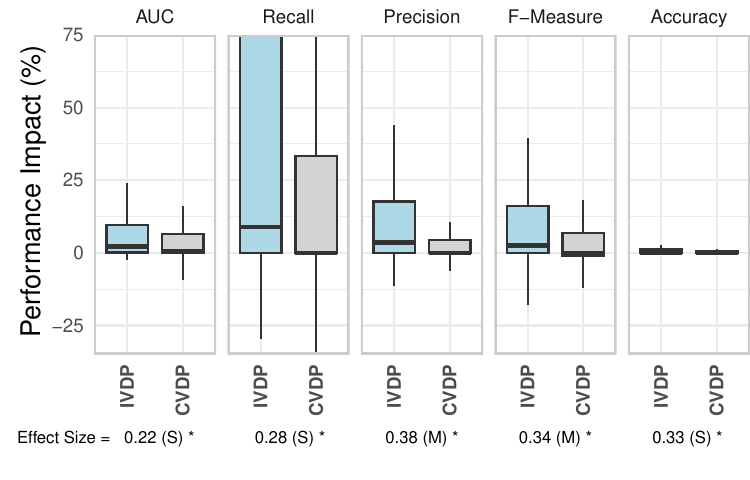}}
 	 
	   \captionsetup[subfloat]{farskip=-2pt,captionskip=-5pt}
	      \captionsetup[subfigure]{labelformat=empty}
	\subfloat[\smaller \textbf{(c) Small Training ( $\leq$ 1/3 Quantile) }]{\includegraphics[width=0.480\linewidth,keepaspectratio]{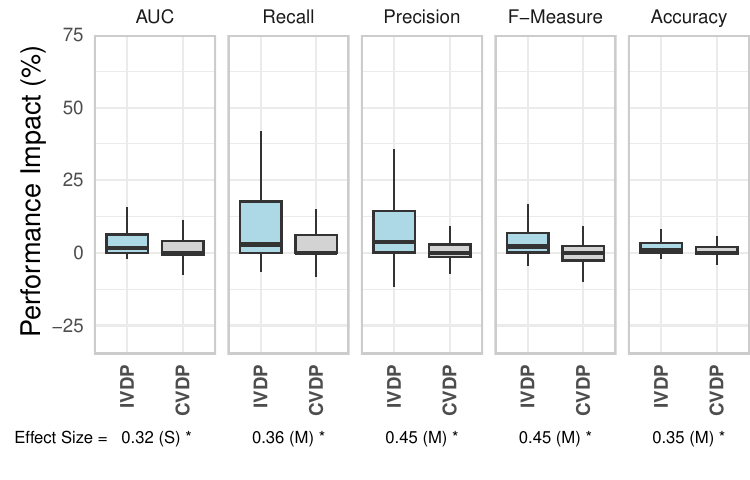}}
	\subfloat[\smaller \textbf{(d) Large Training ( $\ge$ 2/3 Quantile)}]
	{\includegraphics[width=0.480\linewidth, keepaspectratio]{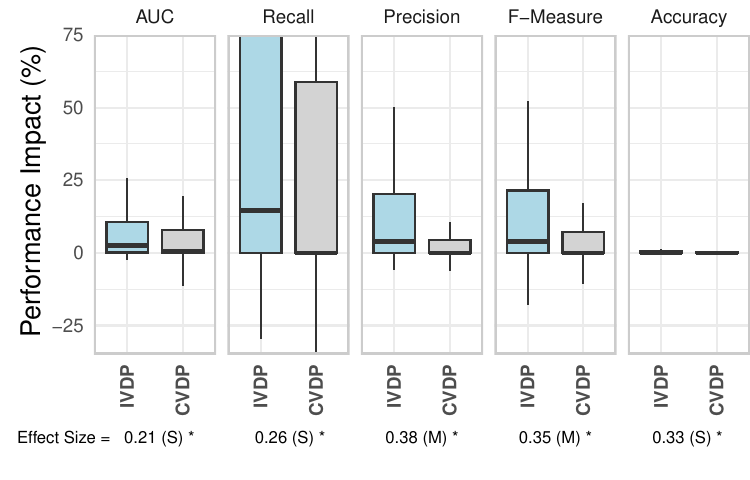}}
\vspace*{-0mm}
	
 	\captionsetup{font={normalsize}}
	\caption{\label{fig:XRQ3Fig1} Distributions of the performance impact $I(M_{\text{tuned}}, M_{\text{default}})$~--- defined in the Equation~(\ref{eq:perImpact1})~--- with grid tuning for the IVDP and CVDP Scenarios.  The $x$-axis on the plots represents the distributions' density. The left panel is small dataset results, while the right panel is large dataset results. The black lines represent the median values of the distributions. The black lines represent the \textbf{median} values of the distributions.  The * represents statistically significant results for the Mann-Whitney U test with $p < 0.05$; $(-)$ denotes $p \ge 0.05$.  N represents \textit{negligible} effect (based on the absolute value of Cliff's delta), S~--- \textit{small} effects, M~--- \textit{medium} effect, and L~--- \textit{large} effect.  For clarity, outliers are removed from the boxplots.} 
\end{figure*}

\noindent \textbf{Results.} Figure~\ref{fig:RQ2CompareAUC} shows the AUC performance impact for each ML model across the studied SDP scenarios. The figure also indicates the significance of the difference and the associated effect size under the associated ML algorithm with the Caret method parameter (see Table~\ref{tab:TableDataset1}).

\noindent \textbf{\textit{The performance gains of many ML algorithms may not hold up across multiple SDP scenarios.  }} For 17 out of 28 studied ML algorithms, Figure~\ref{fig:RQ2CompareAUC} shows that the AUC impact (i.e.,  $I(AUC_{\text{tuned}}, AUC_{\text{default}})$ ) resulted by grid tuning is statistically greater for the IVDP scenario with significant $p$-values and non-negligible effect sizes varying between {\em small} and {\em large}. Interestingly, the difference is negligible for some popular ML implementations (e.g., RF). Moreover, some ML models show a non-negligible difference in performance between the two SDP scenarios. For example, with the grid search tuning, the median AUC impact for MLP model in the IVDP scenario is 26.5\%, whereas it is only 5.9\% for the CVDP scenario.
 Similarly,  15, 24, 18, and 21 out of the total 28 algorithms hold significant differences for the Recall Rate, Precision, F-Measure, and Accuracy optimization metrics, respectively. 
Comparable results hold for the performance impact based on random search tuning. This observed impact of the superiority of hyperparameter tuning in the IVDP scenario raises intriguing questions about the underlying dynamics of the software development life cycle and the selected ML algorithm for SDP. Claiming a significant improvement for some specific ML algorithms under a particular SDP scenario, such as IVDP, may not be valid when getting to more practical scenarios, such as CVDP. A selected ML algorithm for SDP may gain a larger performance impact within a specific software release compared to a scenario where predictions span multiple releases. The results recommend SE researchers and practitioners to consider the significant performance impact variations when selecting an ML algorithm for SDP. Moreover, while these results showcase a consistent trend across various ML algorithms, understanding the relationships between hyperparameter tuning and model architectures for SDP could be an avenue for future exploration.

\noindent \textbf{\textit{The top ML algorithms ranked by the performance impact are common across both SDP scenarios}}. Table~\ref{tab:rq2ranks1} shows the AUC performance impact ranking results from the Scott-Knott test for both SDP scenarios, including all the ML models. For the impact results by grid tuning,  the test groups the 28 ML algorithms into 14 and 9 groups for the IVDP and CVDP scenarios, respectively. We observe that the top ML algorithms ranked by hyperparameter tuning impact are common across the two SDP scenarios, including: {\em mlp}, {\em nnet}, {\em C5.0}, and {\em knn}.  Similar results were obtained for the random-based hyperparameter tuning. The ML algorithms with the lowest ranking in Table~\ref{tab:rq2ranks1} indicate the lowest beneficial performance impact resulting from applying hyperparameter tuning. Unlike other ML algorithms with top impact, it is interesting to notice that powerful Random Forest implementation has the lowest impact for the IVDP scenario and still low rank for the CVDP scenario, with a mean impact from -0.01 to -0.09. Despite the significant difference in performance impact for the majority of ML algorithms across the two SDP scenarios, it is interesting to observe that the top impacted algorithms have some ranking commonality across the SDP scenarios. Similar observations hold for the results by the random algorithm in Table~\ref{tab:rq2ranks1}.

Table~\ref{tab:rq2ranksMLAlgorithhm} presents the actual AUC performance rankings of the ML algorithms before and after grid-based hyperparameter tuning, as determined by the Scott-Knott test. For space reasons, we report the top 15 ML algorithms out of the 28 studied. One key observation when comparing Table~\ref{tab:rq2ranks1} and Table~\ref{tab:rq2ranksMLAlgorithhm} is that the ML algorithms with the highest performance impact due to tuning (from Table~\ref{tab:rq2ranks1}) are not necessarily those that achieve the highest overall performance after tuning (from Table~\ref{tab:rq2ranksMLAlgorithhm}). For instance, in the IVDP scenario, while {\em mlpWeightDecay}, {\em mlp}, {\em nnet}, and {\em C5.0} show the highest tuning impact, they are not the top-performing models after tuning; instead, algorithms like {\em AdaBoost}, {\em Ada}, {\em RF}, and {\em avNNet} rank highest in actual AUC performance.

These results indicate that SE researchers and practitioners can generally expect high-performance gains from the top-ranked ML algorithms in Table~\ref{tab:rq2ranks1} across both SDP scenarios. The presence of consistently top-impacted algorithms, such as {\em mlp} and {\em C5.0}, in both scenarios suggests that these models are particularly responsive to hyperparameter tuning. However, the findings also highlight that a high tuning impact does not always correspond to top overall model performance, as evidenced by the comparison between Tables~\ref{tab:rq2ranks1} and~\ref{tab:rq2ranksMLAlgorithhm}. Ultimately, the choice of ML algorithm plays a key role in determining the impact of tuning, but not necessarily the final performance ranking, regardless of the selected SDP scenario.

\noindent \textbf{\textit{The feature importance differences are inconsistent across the studied SDP scenarios}}. Figure~\ref{fig:XVarImport} presents the feature importance difference distributions across the IVDP and CVDP scenarios. Figure~\ref{fig:XVarImport} presents the distributions of importance differences for the top 30 features (out of a total of 135), ranked by their importance difference values. The remaining features are omitted due to space constraints.  This figure clearly illustrates that there are variations in the feature importance differences between the SDP scenarios, especially for the features from the Promise dataset (see Table~\ref{tab:TableDataset1}).  For example, the features loc, cbo, ce, RFC, and amc are all features from Promise datasets that cover aspects such as the complexity of the software. We observe that feature importance differences for CVDP are wider with larger ranges.   One notable observation is that the maximum difference in feature importance occurs with the CVDP scenario, at an estimated value of 1500, in contrast to 600 in the IVDP scenario. The features in Figure~\ref{fig:XVarImport} are sorted by the maximum importance difference value marked by the red color. Overall, we observe that the rankings are similar between the two SDP scenarios, but there are variations in the rankings for certain features. For instance, the feature $2-cbo$ is ranked second in IVDP while it is ranked fourth in the CVDP scenario. These results indicate that the changes in feature importance due to hyperparameter tuning vary across the SDP scenarios. Future SE researchers and practitioners should consider these differences when analyzing the impacts of tuning across SDP scenarios.

\noindent \textbf{Discussion.} The main scope of this RQ does not focus on which ML algorithm performs best for an SDP scenario before and after hyperparameter tuning. Instead, the focus is on comparing the performance impact of hyperparameter tuning across different SDP scenarios when choosing from a set of ML algorithms. A high-performance impact does not necessarily mean that an ML algorithm is the best performer for a specific SDP scenario and a software project.  For both SDP scenarios, we observed that some ML algorithms show little to no performance gain from hyperparameter tuning. This is often because their default settings already yield performance close to optimal settings for many SE datasets. As a result, even when tuning is applied, the final model performs similarly to the default one, leading to a negligible or low tuning impact. These findings reinforce the idea that hyperparameter tuning is not universally beneficial and that its effectiveness varies depending on possible factors, such as the ML algorithm and SE dataset context~\cite{probst2019tunability}.

While hyperparameter tuning can yield statistically significant performance improvements, these gains must be weighed against the associated computational cost. Our study involved evaluating over 44,000 model configurations across 28 ML algorithms and 53 datasets, requiring substantial computational resources. To help mitigate this cost-performance trade-off, we provide impact-based rankings of ML algorithms (See Table~\ref{tab:rq2ranks1}) across both SDP scenarios. These insights enable researchers and practitioners to prioritize tuning efforts by focusing on algorithms that consistently demonstrate high impact, and potentially skip tuning for those with marginal or inconsistent benefits. As such, our large-scale analysis offers practical guidance for selectively applying tuning where it is most beneficial and cost-effective.

In contrast to prior research, we notice similarities in the most impacted ML algorithms when applying hyperparameter tuning for a particular SDP scenario. For example, the C5.0 algorithm is one of the top models impacted by hyperparameter tuning within the IVDP scenario, as reported by Tantithamthavorn et al$.$~\cite{tantithamthavorn2018impact,tantithamthavorn2016automated}. However, Figure~\ref{fig:RQ2CompareAUC} shows that the impact variation for C5.0 across IVDP and CVDP is not statistically different. That observation is not consistent for all ML algorithms. For instance, the nnet algorithm is ranked as the top impacted ML algorithm in Table~\ref{tab:rq2ranks1}, but the difference across the two scenarios is statistically significant with a medium effect size, as shown in Figure~\ref{fig:RQ2CompareAUC}.  Overall, these results prove variations in the impact across the two studied SDP scenarios from the studied ML algorithms' perspective, unlike the prior research that focuses on individual scenarios~\cite{wahono2014neural,
tantithamthavorn2016automated, fu2016tuning, osman2017hyperparameter, tantithamthavorn2018impact}. The RQ results suggest that the performance impact reported for an ML algorithm in the IVDP scenario does not necessarily apply to other scenarios, such as CVDP.   In addition, while powerful ML models (e.g., Random Forest) may perform well even with the default settings, we cannot conclude that hyperparameter tuning is unnecessary for a given ML algorithm. This is because overall SDP performance can be influenced by multiple factors, such as the choice of features, the number and range of hyperparameters, validation strategy, implementation details of the ML algorithm, and how the data is split, each of which can interact with the tuning process and affect the SDP outcomes~\cite{probst2019tunability}.

Additionally, the results suggest the importance of linking the performance impact and the actual SDP benefits across the frequently used ML algorithms. For example, for a specific software project, a Random Forest based SDP model can still be the top performer despite the low performance impact (see Table~\ref{tab:rq2ranks1}) ranking or the targeted SDP scenario. Moreover, the results suggest that future SE research and practitioners should investigate the relationship between ML algorithms and the expected impact of hyperparameter tuning when evaluating new tuning methods. Future work could further explore the interplay between algorithm characteristics and tuning effects across different SDP scenarios.

\conclusionbox{\textbf{Summary:} The performance impact of hyperparameter tuning varies across ML algorithms and is not consistent between IVDP and CVDP scenarios.
An algorithm with high impact in one scenario (e.g., nnet in CVDP) may not show the same in another (e.g., IVDP), and vice versa.
These results highlight the need to study tuning effects per algorithm and scenario, rather than assuming generalizable behavior. }

\subsection*{RQ3: \textbf{\RQThree}}
\vspace*{-1mm}
\noindent \textbf{Motivation.} 
Understanding how hyperparameter tuning interacts with the size of the software dataset carries crucial implications for the robustness and generalizability of ML models in both studied SDP scenarios. Larger software datasets introduce added complexity, potentially influencing the optimal configuration of hyperparameters and, consequently, the performance impact on ML models. Notably, prior research often selects varying dataset sizes for SDP scenarios~\cite{wahono2014neural,
tantithamthavorn2016automated, fu2016tuning, osman2017hyperparameter, nevendra2022empirical}. The purpose of this study is to determine the relationship between training data size and hyperparameter tuning impact across the studied SDP scenarios. By providing answers to this RQ, practitioners will be guided in understanding possible variations in hyperparameter tuning impacts when using different software training data sizes. Thus, fostering resilient and adaptable tuned SDP methodologies.

\noindent \textbf{Approach.}
In this RQ, we analyze the performance impact changes from the perspective of the studied SE datasets. To derive statistical conclusions, we divide the performance impact results based on dataset training sizes as the number of rows mentioned in Table~\ref{tab:TableDataset1} into two groups of Small and Large. We do the split using two different thresholds. The first threshold is the median (50\% of data), where all data less than or equal to the median is considered a small group, otherwise a large group.  The median measure is robust as outliers do not heavily influence it~\cite{yamashita2014magnet}. The second threshold is applied using the R function \texttt{quantile()}. We use the quantities thresholds 1/3 and 2/3 to determine the number of rows thresholds for the total number of rows to split the performance impacts results per each optimization performance metric (see the 4th Column of Table~\ref{tab:TableDataset1}).  After the median-based split, we have 26 datasets for the Small group and 27 for the Large group. The size of the datasets in the small group ranges between 26 and 723, and in the large group — between 745 and 68993.  After the quantile split, we have 17 datasets for the Small group and 20 for the Large group. The size of the datasets in the small group ranges between 26 and 339, and in the large group — between 1526 and 68993. For small and large dataset groups, we compare the impact of hyperparameter tuning on performance across the IVDP and CVDP scenarios for all the studied performance optimization metrics. To investigate the impact differences, we apply the same statistical tests applied on the previous research questions.

\noindent \textbf{Results.} For IVDP and CVDP scenarios, Figure~\ref{fig:XRQ3Fig1} shows the performance impact $I$ for all the studied performance optimization metrics divided into small and large datasets.

\textbf{\textit{The performance impact has larger effect sizes for small datasets across SDP scenarios}}  In Figure~\ref{fig:XRQ3Fig1}, the results show that both Small and Large datasets have significant differences in the performance impact for both SDP scenarios. In addition, we notice that the difference is noticeably larger for the small training datasets with larger effect sizes. The observation holds for the median-based and quantile-based groups. For example, for the median-based groups, the effect sizes for small datasets range from 0.31 (S) to 0.45 (M) in contrast to a range of 0.19 (S) to 0.37 (M) for large training datasets. The small datasets group has noticeable larger effect sizes for all optimization metrics (see Figures~\ref{fig:XRQ3Fig1}-a and ~\ref{fig:XRQ3Fig1}-c). In contrast, all effect sizes for the large training group are small (see Figures~\ref{fig:XRQ3Fig1}-b and ~\ref{fig:XRQ3Fig1}-d). While both large and small datasets show positive performance gains from hyperparameter tuning within each SDP scenario, the magnitude of the difference in tuning impact between IVDP and CVDP is more clear for the small datasets. In other words, small datasets exhibit higher sensitivity to SDP scenario differences in tuning impact, which results in larger effect sizes. In contrast, larger datasets tend to yield more stable tuning impact across SDP scenarios, which leads to smaller effect sizes when comparing IVDP to CVDP. This finding
complements prior research questions and highlights that the dataset size influences the performance impact difference across the SDP scenarios.

\noindent \textbf{Discussion.} 
When looking deeper into the median performance impacts of each dataset, we notice that the median performance impact per dataset varies across the studied SDP scenarios. For example, for AUC performance impact, the median impact per dataset ranges from 0.37\% to 4.7\% (mean 2.2\%) and -0.40\% to 4.3\% (mean 0.7\%)  for IVDP and CVDP, respectively.  This indicates that differences between the two scenarios happen on the level of the SE datasets. As SE practitioners will usually be interested in a specific SE dataset, seeing the impact variations within the datasets across SDP flags the importance of considering the performance impact differences across SDP scenarios. Overall, these results also indicate that researchers and practitioners should expect more variations in performance impact when leveraging smaller SE datasets. 

One potential explanation for this RQ findings trend is that larger datasets provide more training information, allowing ML algorithms to learn more robust SDP models regardless of whether hyperparameter tuning is applied~\cite{thota2020survey, catal2009investigating}. This makes the relative impact gains from tuning appear more consistent across SPP scenarios. In contrast, smaller datasets are more sensitive to changes in parameter settings, which can amplify the observed differences in tuning impact across SDP scenarios. However, this does not imply that tuning is unnecessary for large datasets. Even powerful ML algorithms may benefit from hyperparameter tuning depending on the dataset characteristics.

The results of this RQ suggest that future research should consider the size of the studied software datasets factor when evaluating the impact of hyperparameter tuning on SDP.  Grouping datasets into small and large subsets can provide deeper insights into performance variations, enabling more focused analyses toward improving performance under different dataset sizes. By evaluating hyperparameter tuning algorithms across both small and large datasets, SE researchers can ensure the robustness and generalizability of their findings, ultimately enhancing the applicability of their studies in SDP scenarios.

\conclusionbox{\textbf{Summary:} The impact of hyperparameter tuning varies with software dataset size, with smaller datasets showing greater variability across SDP scenarios.
Performance differences between IVDP and CVDP often emerge at the dataset level, not just the scenario level.
SE Researchers should consider dataset size when evaluating tuning impact to ensure robust and generalizable SDP findings across scenarios.  }

 \section{Implications}\label{sec:implications}
In this section, we outline the implications of our results, including potential pitfalls to avoid, recommendations, and future research opportunities.

\textit{\textbf{Implication (1)}: The SE researchers and practitioners should be aware
that the performance impact of hyperparameter tuning could be dependent on the selected SDP scenario. We recommend that SE researchers and practitioners consider the variation in the performance impact when evaluating a new tuning method or deploying a production model}. From RQ1 results, we find that the performance impact can be significantly different across SDP scenarios. In other words, evaluating and reporting a performance impact on an SDP scenario like IVDP can be significantly misleading for future researchers and practitioners. We encourage future researchers to explore the impact on at least two SDP scenarios to give a wider insight into the hyperparameter tuning impact.  

Furthermore, we recommend future research to investigate further the negative performance
impacts of applying hyperparameter tuning techniques on SDP performance, as SE practitioners may be misled by focusing solely on the positive impact of tuning hyperparameters. Negative impacts of hyperparameter tuning, particularly in CVDP scenarios, may arise due to data distribution shifts between training and testing releases. In CVDP, models are trained and tuned on past versions, but tested on future releases where the characteristics of the data (e.g., feature distributions or defect densities) may differ significantly~\cite{gangwar2023concept, kabir2019assessing}. Future SE researchers can build upon RQ1 findings and develop adaptive or robust tuning strategies that mitigate these effects in practical SDP deployments.

\textit{\textbf{Implication (2)}: SE  researchers and practitioners should consider the importance of linking the performance impact and the actual SDP benefits across the
frequently used ML algorithms.} From the results presented in RQ2, we observe that the performance impact varies significantly for some ML algorithms across SDP scenarios. High-performance impact does not necessarily indicate that an ML algorithm is the best performer for a specific SDP scenario and software project. It may simply indicate that the ML algorithm benefits significantly from being fine-tuned for that particular SDP scenario. The results highlight the importance of linking performance impact with the actual SDP benefits across commonly used ML algorithms. For instance, a Random Forest based SDP model can still be the top performer for a particular software project, even if it ranks low in performance impact (see Tables~\ref{tab:rq2ranks1} and ~\ref {tab:rq2ranksMLAlgorithhm}) or for the specific SDP scenario. Furthermore, the findings suggest that future SE research and practitioners should explore the relationship between applied ML algorithms and the expected impact of hyperparameter tuning when evaluating new tuning methods. 

In that relationship, we expect the research to highlight the importance of evaluating both the relative performance impact (i.e., percentage of performance impact after tuning) and the absolute performance level (i.e., Actual Performance Values) of each ML algorithm. Future SE research should aim to systematically analyze how specific ML algorithms respond to tuning across  SDP scenarios. By doing so, we can better understand which algorithms are inherently robust (i.e., perform well with default settings), which are tuning-sensitive, and under what conditions tuning is most beneficial. These insights can help practitioners make informed choices when selecting ML algorithms and deciding whether tuning is worth the additional computational cost in their SDP scenario.

\textit{\textbf{Implication (3)}: When considering the impact of hyperparameter tuning, the project complexity should be considered across the SDP scenarios.} The impact of hyperparameter tuning is significantly different across SDP when considering different training datasets. SE researchers and practitioners should take into consideration the size of datasets when targeting generalizable findings for new tuning algorithms or selected ML models.

 Based on the study findings, we can highlight a set of recommendations to help SE practitioners accelerate the decision-making process when working with hyperparameter tuning:

\begin{itemize}[leftmargin=*,label={}]
	\item  \textbf{SDP Scenario-Aware Tuning:}   Practitioners should first identify whether their SDP task aligns more with an IVDP or a CVDP setup. Since IVDP involves tuning and testing within the same version, our results show that tuning generally yields more consistent improvements. Thus, tuning is recommended and likely beneficial in IVDP, given its relatively low risk. In contrast, CVDP shows a higher chance of negative or marginal improvements. 
 
 \item  \textbf{ML Algorithm-Tuning Sensitivity:}  Our results reveal that some ML algorithms (e.g., C5.0 and mlp) are more sensitive to tuning than others across SDP scenarios. For practitioners, this means that instead of testing all ML algorithms, one could prioritize ML algorithms known to benefit more consistently across SDP scenarios, while using default configurations for less sensitive ones (e.g., Random Forest).

\item  \textbf{Dataset Size Consideration:}  As also discussed in RQ3, the effectiveness of tuning varies with dataset size across SDP scenarios. For smaller SE datasets, practitioners could expect larger differences in performance across SDPs and spend more effort exploring the impact of tuning.  This allows practitioners to determine whether tuning effort is justified given the availability of training data.

\item  \textbf{Avoiding Brute-force Search:}   Based on our findings, we do not recommend a brute-force search over all ML algorithms and tuning settings. Instead, developers may choose a subset of representative ML algorithms, tuned selectively depending on the SDP scenarios. Also, using  default parameters when the expected tuning impact is low across SDP scenarios could be a good option. 

\end{itemize}

Overall, these recommendations include  possible actions that could help accelerate the decision-making process by aligning tuning efforts with the SDP scenarios, dataset size, and model sensitivity, reducing unnecessary computation while still achieving strong SDP  performance.

\section{threats}\label{sec:threats}

In this section, we discuss the threats to the validity of this study work, classified as per \cite{wohlin2012experimentation,yin2009case}.

\noindent\textbf{Construct Validity.} We analyze several collections of
datasets (e.g., NASA and PROMISE). Each collection of datasets has different
sets of metrics. This variation of metrics could impact our results.
Nevertheless, prior research from Tantithamthavorn et
al.~\cite{tantithamthavorn2018impact} has shown
that the number and types of metrics within these collections of datasets do
not significantly influence the findings derived from them. 

Our analyses of low  Events Per Variable (EPV)~\cite{tantithamthavorn2016empirical} datasets pose another threat. Low EPV datasets may produce models
with high overfitting risks, which could cause our analyses on these models to
be unfruitful. Nevertheless, it is important to analyze low EPV datasets, as in industrial settings, datasets with a low EPV may be the only option for some
software projects. In such cases, taking the risk of
overfitting a model may be the only choice. Moreover, the fact that low EPV
datasets are the majority in previous research (e.g., 77\% of the datasets have
a low EPV in the previous work performed by Tantithamthavorn et
al.~\cite{tantithamthavorn2016automated,tantithamthavorn2018impact}), shows the
importance of analyzing such low EPV datasets.   

In the evaluation of SDP scenarios,  we applied bootstrap validation for model training and validation. We recognize that other splitting validation methods could be used
instead. For example, one could use 10-fold cross-validation to build the training
and validation sets. In future studies, we recommend exploring the impact of more advanced splitting mechanisms. Finally, we use the grid search and random search in this work. Many other alternative algorithms are available, such as genetic algorithm and differential evolution.
Although previous research has shown that the grid search algorithm yield
performance improvements similar to the aforementioned
alternatives~\cite{tantithamthavorn2018impact}, we recommend future research to investigate the generalizability of this study's results to other hyperparameter tuning algorithms. 

According to Section~\ref{subsec:edm}, 44,520 experiments must be performed. A fraction of these experiments, however, failed due to hardware limitations and software instability. In order to combat this threat, we developed a software harness that re-ran the failing jobs, helping to successfully complete $\approx$~94\% of all of the jobs (i.e., 41,741 out of 44,520). 
As 94\% of the jobs are completed successfully, the conclusions are not significantly affected by the remaining failing jobs.

In RQ3, we define small and large datasets using median and quantile-based thresholds. While these approaches are commonly used in empirical SE research~\cite{rakha2016studying, catal2009investigating}, we acknowledge that the boundary between small and large is context-dependent.  This may influence the observed effect sizes and should be interpreted as relative to the sample datasets used in this study. Future work could explore alternative grouping strategies to confirm the generalizability of our findings.

\noindent\textbf{Internal Validity.} We evaluate the performance impact of hyperparameter tuning on the studied ML models
using AUC, recall rate, precision, F-measure, and accuracy scores. However, other measures may yield different results. We recommend future studies to expand the set of performance measures to include other metrics, such as Mean Square Error. Previous work has
shown that noisy data may taunt the conclusions that are drawn from defect
prediction
studies~\cite{ghotra2015revisiting,tantithamthavorn2016towards,tantithamthavorn2015impact}.
Therefore, known-to-be noisy NASA data~\cite{shepperd2013data} could be
influencing our conclusions. To combat this threat, we rely on cleaned datasets
provided by Tantithamthavorn~\cite{defectDataR2019}. 

While prior research has established bootstrap validation as standard in software defect prediction (SDP)~\cite{gong2021revisiting, tantithamthavorn2016towards}, using a uniform bootstrap process across all ML algorithms may not fully capture their unique characteristics. Although this method effectively balances bias and variance, tailored validation could yield better insights for specific algorithms. We chose a uniform approach for consistency and comparability, and to mitigate this limitation, we kept other factors, such as data splits and random seeds, constant to isolate the impact of hyperparameter tuning.

While our study adopts tuning ranges from prior work where available~\cite{fu2016tuning, tantithamthavorn2016towards}, and uses a fixed  \texttt{Caret}  \texttt{tuneLength} of $5$ otherwise, we acknowledge that the choice of search space boundaries and granularity (i.e., number of values per parameter) can influence the tuning results. Automatically generated search spaces in  \texttt{Caret} are based on model-specific heuristics, which may not capture all relevant parameter ranges for every dataset or algorithm. Furthermore, fixing  \texttt{tuneLength} to 5 may limit the exploration of the whole hyperparameter space for complex models, potentially affecting the observed performance gains. However, this decision was made to ensure consistent and computationally feasible comparisons across 28 algorithms and 53 post-release datasets. Future work can explore the impact of larger or data-driven search spaces on tuning effectiveness~\cite{ali2023hyperparameter}.

The selection of an IVDP experiment setup could raise the potential risk of overestimating model performance due to the lack of unseen test data. However, in our study, we apply the bootstrap validation method during the selection of hyperparameters. In each iteration of the bootstrap process, 10\% of the data is set aside as test data to evaluate the hyperparameter settings. This ensures that hyperparameters are evaluated on data that has not been used in training, which helps mitigate the issue of overfitting. Additionally, the performance impact of IVDP in our study is assessed for both tuned and untuned settings using a consistent performance calculation method. The IVDP scenario, while different from CVDP, is commonly used in hyperparameter tuning research due to its practicality in certain SDP contexts~\cite{gong2021revisiting, tantithamthavorn2016towards, tantithamthavorn2016empirical}.

\noindent\textbf{External Validity.} 
We study 53 post-release datasets in this paper. Despite studying more datasets than previous research~\cite{fu2016tuning,khan2020hyper, nevendra2022empirical,tantithamthavorn2018impact}, our results are still limited to the characteristics of the selected datasets and may not generalize to all software systems. In particular, while we included several new datasets to broaden the evaluation, our dataset pool still leans toward widely used open-source projects, which may introduce a dataset bias. This bias may limit the applicability of our findings to systems with different development practices, domains, or scales. According to Wieringa et al.~\cite{26c990771bd645428c33ea107259ceb5}, building a theory requires generalization to a broader theoretical population and understanding the architectural similarity of the studied systems. Nevertheless, our goal is not to demonstrate universally generalized results but to assess the relative performance impact of hyperparameter tuning across SDP scenarios. Using well-designed and controlled experiments, the same empirical methodology can be applied to other contexts and scenarios (e.g., cross-project or industrial SDP) in future work.

This study focuses exclusively on traditional ML algorithms applied to structured, tabular defect datasets. While these models remain widely used in practice~\cite{stradowski2023industrial} and are well-aligned with the feature-based datasets leveraged in this study, we acknowledge that our findings may not generalize to settings where unstructured data (e.g., source code or textual descriptions) are available or where large pretrained models, such as Large Language Models (LLMs)~\cite{gao2025current}, are applicable. As pretrained and neural models may follow different performance behaviors and tuning sensitivities, future research is needed to assess whether the conclusions observed in this study hold when applied to such architectures.

\section{Conclusion}\label{sec:conclusion}

In this paper, we contrast the impact of hyperparameter tuning across two primary SDP scenarios. We set out to study 53 post-release datasets from
the PROMISE~\cite{he2012investigation} and NASA~\cite{shepperd2013data} repositories, as well as datasets provided by Kim et
al.~\cite{kim2011dealing}, D'Ambros et al.~\cite{d2010extensive}, Zimmermann et
al.~\cite{zimmermann2007predicting}. The study investigates
how the impact of hyperparameter tuning could vary across different SDP scenarios. To
this end, we evaluate a list of 28 ML algorithms using tuned and default settings across two pivotal and consecutive two SDP scenarios, namely, IVDP and CVDP.    We also investigate the impact of two widely used tuning algorithms, Grid, and Random Search algorithms. Finally, we evaluate the performance impact using a set of performance metrics such as AUC and recall rate. 

Results include the following main observations. 
\begin{enumerate*}
    \item  The IVDP scenario gains a larger performance improvement than the CVDP scenario when hyperparameters are tuned for SDP. The difference is statistically significant with non-negligible effect sizes for four optimization metrics. 
    \item The performance gains of many ML algorithms may not hold up across multiple SDP scenarios.  This difference is widespread: up to 22 out of 28 (78\%) applied ML algorithms show statistically significant differences in performance impacts for SDP.
    \item The small software datasets are affected more than the large ones by the differences in the performance impacts. 
\end{enumerate*}

The primary novelty focus of this study is to compare the performance impact of hyperparameter tuning across different SDP scenarios. Unlike prior research, which has typically examined the impact of hyperparameter tuning within a single SDP scenario using various approaches and scales, our study addresses the diversity found in separate studies with different experimental designs and SDP scenarios. We aim to demonstrate the variations in hyperparameter tuning impact across multiple SDP scenarios. Moreover, our comparative analysis is conducted on a large scale, encompassing the widest range of ML algorithms and the largest datasets studied to date. These aspects underscore the unique contributions of this research. 

Given the computationally hungry nature of our analyses, we focused on contrasting the performance impact across SDP scenarios using Grid and Random tuning. Considerable computational complexity in our study also covers a large number of datasets and ML algorithms across two SDP scenarios. However,  exploring additional tuning methods could give further insights, and we will consider it for future work.

In conclusion, this study highlights an important perspective when it comes to studying the impact of hyperparameter tuning methods on an SDP challenge. The study suggests that future studies should pay attention to the variations in the performance impact resulting from hyperparameter tuning across different SDP scenarios. 
Additionally, the results recommend researchers to link the expected performance impact to the ML algorithm used and the SDP scenario studied.

\balance
\bibliographystyle{IEEEtran}
\bibliography{RefBio}

\begin{thebibliography}{10}
\providecommand{\url}[1]{#1}
\csname url@samestyle\endcsname
\providecommand{\newblock}{\relax}
\providecommand{\bibinfo}[2]{#2}
\providecommand{\BIBentrySTDinterwordspacing}{\spaceskip=0pt\relax}
\providecommand{\BIBentryALTinterwordstretchfactor}{4}
\providecommand{\BIBentryALTinterwordspacing}{\spaceskip=\fontdimen2\font plus
\BIBentryALTinterwordstretchfactor\fontdimen3\font minus
  \fontdimen4\font\relax}
\providecommand{\BIBforeignlanguage}[2]{{%
\expandafter\ifx\csname l@#1\endcsname\relax
\typeout{** WARNING: IEEEtran.bst: No hyphenation pattern has been}%
\typeout{** loaded for the language `#1'. Using the pattern for}%
\typeout{** the default language instead.}%
\else
\language=\csname l@#1\endcsname
\fi
#2}}
\providecommand{\BIBdecl}{\relax}
\BIBdecl

\bibitem{zimmermann2007predicting}
T.~Zimmermann, R.~Premraj, and A.~Zeller, ``Predicting defects for eclipse,''
  in \emph{Predictor Models in Software Engineering, 2007. PROMISE'07: ICSE
  Workshops 2007. International Workshop on}.\hskip 1em plus 0.5em minus
  0.4em\relax IEEE, 2007, pp. 9--9.

\bibitem{wahono2014neural}
R.~S. Wahono, N.~S. Herman, and S.~Ahmad, ``Neural network parameter
  optimization based on genetic algorithm for software defect prediction,''
  \emph{Advanced Science Letters}, vol.~20, no. 10-11, pp. 1951--1955, 2014.

\bibitem{bowes2015different}
D.~Bowes, T.~Hall, and J.~Petri{\'c}, ``Different classifiers find different
  defects although with different level of consistency,'' in \emph{Proc. of the
  11th International Conference on Predictive Models and Data Analytics in
  Software Engineering}.\hskip 1em plus 0.5em minus 0.4em\relax ACM, 2015,
  p.~3.

\bibitem{d2010extensive}
M.~D'Ambros, M.~Lanza, and R.~Robbes, ``An extensive comparison of bug
  prediction approaches,'' in \emph{Mining Software Repositories (MSR), 2010
  7th IEEE Working Conference on}.\hskip 1em plus 0.5em minus 0.4em\relax IEEE,
  2010, pp. 31--41.

\bibitem{kim2011dealing}
S.~Kim, H.~Zhang, R.~Wu, and L.~Gong, ``Dealing with noise in defect
  prediction,'' in \emph{Software Engineering (ICSE), 2011 33rd International
  Conference on}.\hskip 1em plus 0.5em minus 0.4em\relax IEEE, 2011, pp.
  481--490.

\bibitem{he2012investigation}
Z.~He, F.~Shu, Y.~Yang, M.~Li, and Q.~Wang, ``An investigation on the
  feasibility of cross-project defect prediction,'' \emph{Automated Software
  Engineering}, vol.~19, no.~2, pp. 167--199, 2012.

\bibitem{zhang2014towards}
F.~Zhang, A.~Mockus, I.~Keivanloo, and Y.~Zou, ``Towards building a universal
  defect prediction model,'' in \emph{Proc. of the Working Conference on Mining
  Software Repositories}.\hskip 1em plus 0.5em minus 0.4em\relax ACM, 2014, pp.
  182--191.

\bibitem{zhang2016cross}
F.~Zhang, Q.~Zheng, Y.~Zou, and A.~E. Hassan, ``Cross-project defect prediction
  using a connectivity-based unsupervised classifier,'' in \emph{Proc. of the
  38th International Conference on Software Engineering}.\hskip 1em plus 0.5em
  minus 0.4em\relax ACM, 2016, pp. 309--320.

\bibitem{gong2021revisiting}
L.~Gong, G.~K. Rajbahadur, A.~E. Hassan, and S.~Jiang, ``Revisiting the impact
  of dependency network metrics on software defect prediction,'' \emph{IEEE
  Transactions on Software Engineering}, vol.~48, no.~12, pp. 5030--5049, 2021.

\bibitem{tantithamthavorn2016automated}
C.~Tantithamthavorn, S.~McIntosh, A.~E. Hassan, and K.~Matsumoto, ``Automated
  parameter optimization of classification techniques for defect prediction
  models,'' in \emph{Software Engineering (ICSE), 2016 IEEE/ACM 38th
  International Conference on}.\hskip 1em plus 0.5em minus 0.4em\relax IEEE,
  2016, pp. 321--332.

\bibitem{fu2016tuning}
W.~Fu, T.~Menzies, and X.~Shen, ``Tuning for software analytics: Is it really
  necessary?'' \emph{Information and Software Technology}, vol.~76, pp.
  135--146, 2016.

\bibitem{osman2017hyperparameter}
H.~Osman, M.~Ghafari, and O.~Nierstrasz, ``Hyperparameter optimization to
  improve bug prediction accuracy,'' in \emph{2017 IEEE Workshop on Machine
  Learning Techniques for Software Quality Evaluation (MaLTeSQuE)}.\hskip 1em
  plus 0.5em minus 0.4em\relax IEEE, 2017, pp. 33--38.

\bibitem{tantithamthavorn2018impact}
C.~Tantithamthavorn, S.~McIntosh, A.~E. Hassan, and K.~Matsumoto, ``The impact
  of automated parameter optimization on defect prediction models,'' \emph{IEEE
  Transactions on Software Engineering}, 2018.

\bibitem{XcaretPkg2019}
M.~Kuhn, ``Caret package,''
  \url{https://cran.r-project.org/web/packages/caret/index.html}, 2019.

\bibitem{WekaSearchPackage}
``Multisearch-weka-package,''
  \url{https://github.com/fracpete/multisearch-weka-package}, 2020.

\bibitem{omran2005differential}
M.~G. Omran, A.~P. Engelbrecht, and A.~Salman, ``Differential evolution methods
  for unsupervised image classification,'' in \emph{Evolutionary Computation,
  2005. The 2005 IEEE Congress on}, vol.~2.\hskip 1em plus 0.5em minus
  0.4em\relax IEEE, 2005, pp. 966--973.

\bibitem{kondo2019impact}
M.~Kondo, C.-P. Bezemer, Y.~Kamei, A.~E. Hassan, and O.~Mizuno, ``The impact of
  feature reduction techniques on defect prediction models,'' \emph{Empirical
  Software Engineering}, pp. 1--39, 2019.

\bibitem{li2020understanding}
K.~Li, Z.~Xiang, T.~Chen, S.~Wang, and K.~C. Tan, ``Understanding the automated
  parameter optimization on transfer learning for cpdp: An empirical study,''
  \emph{arXiv preprint arXiv:2002.03148}, 2020.

\bibitem{rajbahadur2019impact}
G.~K. Rajbahadur, S.~Wang, Y.~Kamei, and A.~E. Hassan, ``Impact of
  discretization noise of the dependent variable on machine learning
  classifiers in software engineering,'' \emph{IEEE Transactions on Software
  Engineering}, 2019.

\bibitem{agrawal2018better}
A.~Agrawal and T.~Menzies, ``Is" better data" better than" better data miners"?
  on the benefits of tuning smote for defect prediction,'' in \emph{Proc. of
  the 40th International Conference on Software engineering}, 2018, pp.
  1050--1061.

\bibitem{ali2021empirical}
A.~Ali and C.~Gravino, ``An empirical comparison of validation methods for
  software prediction models,'' \emph{Journal of Software: Evolution and
  Process}, vol.~33, no.~8, p. e2367, 2021.

\bibitem{fu2016differential}
W.~Fu, V.~Nair, and T.~Menzies, ``Why is differential evolution better than
  grid search for tuning defect predictors?'' \emph{arXiv preprint
  arXiv:1609.02613}, 2016.

\bibitem{lee2022holistic}
J.~Lee, J.~Choi, D.~Ryu, and S.~Kim, ``Holistic parameter optimization for
  software defect prediction,'' \emph{IEEE Access}, vol.~10, pp.
  106\,781--106\,797, 2022.

\bibitem{agrawal2021simpler}
A.~Agrawal, X.~Yang, R.~Agrawal, R.~Yedida, X.~Shen, and T.~Menzies, ``Simpler
  hyperparameter optimization for software analytics: Why, how, when?''
  \emph{IEEE Transactions on Software Engineering}, vol.~48, no.~8, pp.
  2939--2954, 2021.

\bibitem{shrikanth2021early}
N.~Shrikanth, S.~Majumder, and T.~Menzies, ``Early life cycle software defect
  prediction. why? how?'' in \emph{2021 IEEE/ACM 43rd International Conference
  on Software Engineering (ICSE)}.\hskip 1em plus 0.5em minus 0.4em\relax IEEE,
  2021, pp. 448--459.

\bibitem{duan2003evaluation}
K.~Duan, S.~S. Keerthi, and A.~N. Poo, ``Evaluation of simple performance
  measures for tuning svm hyperparameters,'' \emph{Neurocomputing}, vol.~51,
  pp. 41--59, 2003.

\bibitem{probst2019tunability}
P.~Probst, A.-L. Boulesteix, and B.~Bischl, ``Tunability: Importance of
  hyperparameters of machine learning algorithms,'' \emph{Journal of Machine
  Learning Research}, vol.~20, no.~53, pp. 1--32, 2019.

\bibitem{eiben2011parameter}
A.~E. Eiben and S.~K. Smit, ``Parameter tuning for configuring and analyzing
  evolutionary algorithms,'' \emph{Swarm and Evolutionary Computation}, vol.~1,
  no.~1, pp. 19--31, 2011.

\bibitem{rajbahadur2021impact}
G.~K. Rajbahadur, S.~Wang, G.~A. Oliva, Y.~Kamei, and A.~E. Hassan, ``The
  impact of feature importance methods on the interpretation of defect
  classifiers,'' \emph{IEEE Transactions on Software Engineering}, vol.~48,
  no.~7, pp. 2245--2261, 2021.

\bibitem{jiarpakdee2021practitioners}
J.~Jiarpakdee, C.~K. Tantithamthavorn, and J.~Grundy, ``Practitioners’
  perceptions of the goals and visual explanations of defect prediction
  models,'' in \emph{2021 IEEE/ACM 18th International Conference on Mining
  Software Repositories (MSR)}.\hskip 1em plus 0.5em minus 0.4em\relax IEEE,
  2021, pp. 432--443.

\bibitem{back1996evolutionary}
T.~Back, \emph{Evolutionary algorithms in theory and practice: evolution
  strategies, evolutionary programming, genetic algorithms}.\hskip 1em plus
  0.5em minus 0.4em\relax Oxford university press, 1996.

\bibitem{diaz1983method}
A.~R. Diaz, N.~Kikuchi, and J.~E. Taylor, ``A method of grid optimization for
  finite element methods,'' \emph{Computer Methods in Applied Mechanics and
  Engineering}, vol.~41, no.~1, pp. 29--45, 1983.

\bibitem{bergstra2012random}
J.~Bergstra and Y.~Bengio, ``Random search for hyper-parameter optimization,''
  \emph{Journal of Machine Learning Research}, vol.~13, no. Feb, pp. 281--305,
  2012.

\bibitem{ho1995random}
T.~K. Ho, ``Random decision forests,'' in \emph{Proc. of 3rd international
  conference on document analysis and recognition}, vol.~1.\hskip 1em plus
  0.5em minus 0.4em\relax IEEE, 1995, pp. 278--282.

\bibitem{tantithamthavorn2016empirical}
C.~Tantithamthavorn, S.~McIntosh, A.~E. Hassan, and K.~Matsumoto, ``An
  empirical comparison of model validation techniques for defect prediction
  models,'' \emph{IEEE Transactions on Software Engineering}, vol.~43, no.~1,
  pp. 1--18, 2016.

\bibitem{khan2020hyper}
F.~Khan, S.~Kanwal, S.~Alamri, and B.~Mumtaz, ``Hyper-parameter optimization of
  classifiers, using an artificial immune network and its application to
  software bug prediction,'' \emph{Ieee Access}, vol.~8, pp. 20\,954--20\,964,
  2020.

\bibitem{nevendra2022empirical}
M.~Nevendra and P.~Singh, ``Empirical investigation of hyperparameter
  optimization for software defect count prediction,'' \emph{Expert Systems
  with Applications}, vol. 191, p. 116217, 2022.

\bibitem{gray2012reflections}
D.~Gray, D.~Bowes, N.~Davey, Y.~Sun, and B.~Christianson, ``Reflections on the
  nasa mdp data sets,'' \emph{IET software}, vol.~6, no.~6, pp. 549--558, 2012.

\bibitem{defectDataR2019}
C.~Tantithamthavorn, ``R package defectdata,''
  \url{https://github.com/awsm-research/DefectData}, 2016.

\bibitem{xu2019tstss}
Z.~Xu, S.~Li, X.~Luo, J.~Liu, T.~Zhang, Y.~Tang, J.~Xu, P.~Yuan, and J.~Keung,
  ``Tstss: A two-stage training subset selection framework for cross version
  defect prediction,'' \emph{Journal of Systems and Software}, vol. 154, pp.
  59--78, 2019.

\bibitem{sarro2012further}
F.~Sarro, S.~Di~Martino, F.~Ferrucci, and C.~Gravino, ``A further analysis on
  the use of genetic algorithm to configure support vector machines for
  inter-release fault prediction,'' in \emph{Proc. of the 27th annual ACM
  symposium on applied computing}, 2012, pp. 1215--1220.

\bibitem{nikravesh2023parameter}
N.~Nikravesh and M.~R. Keyvanpour, ``Parameter tuning for software fault
  prediction with different variants of differential evolution,'' \emph{Expert
  Systems with Applications}, p. 121251, 2023.

\bibitem{rajbahadur2017impact}
G.~K. Rajbahadur, S.~Wang, Y.~Kamei, and A.~E. Hassan, ``The impact of using
  regression models to build defect classifiers,'' in \emph{Proc. of the 14th
  International Conference on Mining Software Repositories}.\hskip 1em plus
  0.5em minus 0.4em\relax IEEE Press, 2017, pp. 135--145.

\bibitem{ghotra2015revisiting}
B.~Ghotra, S.~McIntosh, and A.~E. Hassan, ``Revisiting the impact of
  classification techniques on the performance of defect prediction models,''
  in \emph{Proc. of the 37th International Conference on Software
  Engineering-Volume 1}.\hskip 1em plus 0.5em minus 0.4em\relax IEEE Press,
  2015, pp. 789--800.

\bibitem{shepperd2013data}
M.~Shepperd, Q.~Song, Z.~Sun, and C.~Mair, ``Data quality: Some comments on the
  nasa software defect datasets,'' \emph{IEEE Transactions on Software
  Engineering}, vol.~39, no.~9, pp. 1208--1215, 2013.

\bibitem{falessi2023enhancing}
D.~Falessi, S.~M. Laureani, J.~{\c{C}}arka, M.~Esposito, and D.~A.~d. Costa,
  ``Enhancing the defectiveness prediction of methods and classes via jit,''
  \emph{Empirical Software Engineering}, vol.~28, no.~2, p.~37, 2023.

\bibitem{rakha2016studying}
M.~S. Rakha, W.~Shang, and A.~E. Hassan, ``Studying the needed effort for
  identifying duplicate issues,'' \emph{Empirical Software Engineering},
  vol.~21, no.~5, pp. 1960--1989, 2016.

\bibitem{RpackageCite}
\BIBentryALTinterwordspacing
{R Core Team}, \emph{R: A Language and Environment for Statistical Computing, R
  version 4.3.1 (2023-06-16)}, R Foundation for Statistical Computing, Vienna,
  Austria, 2023. [Online]. Available: \url{https://www.R-project.org/}
\BIBentrySTDinterwordspacing

\bibitem{RcaretCite}
\BIBentryALTinterwordspacing
M.~Kuhn, ``Building predictive models in r using the caret package - version
  6.0.94,'' \emph{Journal of Statistical Software}, vol.~28, no.~5, p. 1–26,
  2008. [Online]. Available:
  \url{https://www.jstatsoft.org/index.php/jss/article/view/v028i05}
\BIBentrySTDinterwordspacing

\bibitem{boughorbel2017optimal}
S.~Boughorbel, F.~Jarray, and M.~El-Anbari, ``Optimal classifier for imbalanced
  data using matthews correlation coefficient metric,'' \emph{PloS one},
  vol.~12, no.~6, p. e0177678, 2017.

\bibitem{lessmann2008benchmarking}
S.~Lessmann, B.~Baesens, C.~Mues, and S.~Pietsch, ``Benchmarking classification
  models for software defect prediction: A proposed framework and novel
  findings,'' \emph{IEEE Transactions on Software Engineering}, vol.~34, no.~4,
  pp. 485--496, 2008.

\bibitem{efron1983estimating}
B.~Efron, ``Estimating the error rate of a prediction rule: improvement on
  cross-validation,'' \emph{Journal of the American statistical association},
  vol.~78, no. 382, pp. 316--331, 1983.

\bibitem{yatish2019mining}
S.~Yatish, J.~Jiarpakdee, P.~Thongtanunam, and C.~Tantithamthavorn, ``Mining
  software defects: Should we consider affected releases?'' in \emph{2019
  IEEE/ACM 41st International Conference on Software Engineering (ICSE)}.\hskip
  1em plus 0.5em minus 0.4em\relax IEEE, 2019, pp. 654--665.

\bibitem{varian2005bootstrap}
H.~Varian, ``Bootstrap tutorial,'' \emph{Mathematica Journal}, vol.~9, no.~4,
  pp. 768--775, 2005.

\bibitem{CCDB}
``Digital research alliance of canada,'' \url{https://alliancecan.ca}, 2023.

\bibitem{Gehan1965}
E.~A. Gehan, ``A generalized {W}ilcoxon test for comparing arbitrarily
  singly-censored samples,'' \emph{Biometrika}, vol.~52, no. 1-2, pp. 203--223,
  1965.

\bibitem{Cliff1}
J.~D. Long, D.~Feng, and N.~Cliff, ``Ordinal analysis of behavioral data,''
  \emph{Handbook of psychology}, 2003.

\bibitem{Cliff2}
J.~Romano, J.~D. Kromrey, J.~Coraggio, J.~Skowronek, and L.~Devine, ``Exploring
  methods for evaluating group differences on the nsse and other surveys: Are
  the t-test and {C}ohens'd indices the most appropriate choices,'' in
  \emph{annual meeting of the Southern Association for Institutional Research},
  2006.

\bibitem{hart2001mann}
A.~Hart, ``Mann-whitney test is not just a test of medians: differences in
  spread can be important,'' \emph{Bmj}, vol. 323, no. 7309, pp. 391--393,
  2001.

\bibitem{gangwar2023concept}
A.~K. Gangwar and S.~Kumar, ``Concept drift in software defect prediction: A
  method for detecting and handling the drift,'' \emph{ACM Transactions on
  Internet Technology}, vol.~23, no.~2, pp. 1--28, 2023.

\bibitem{kabir2019assessing}
M.~A. Kabir, J.~W. Keung, K.~E. Bennin, and M.~Zhang, ``Assessing the
  significant impact of concept drift in software defect prediction,'' in
  \emph{2019 IEEE 43rd Annual Computer Software and Applications Conference
  (COMPSAC)}, vol.~1.\hskip 1em plus 0.5em minus 0.4em\relax IEEE, 2019, pp.
  53--58.

\bibitem{tantithamthavorn2015impact}
C.~Tantithamthavorn, S.~McIntosh, A.~E. Hassan, A.~Ihara, and K.~Matsumoto,
  ``The impact of mislabelling on the performance and interpretation of defect
  prediction models,'' in \emph{IEEE/ACM IEEE International Conference on
  Software Engineering}, vol.~1.\hskip 1em plus 0.5em minus 0.4em\relax IEEE,
  2015, pp. 812--823.

\bibitem{yamashita2014magnet}
K.~Yamashita, S.~McIntosh, Y.~Kamei, and N.~Ubayashi, ``Magnet or sticky? an
  oss project-by-project typology,'' in \emph{Proceedings of the 11th working
  conference on mining software repositories}, 2014, pp. 344--347.

\bibitem{thota2020survey}
M.~K. Thota, F.~H. Shajin, P.~Rajesh \emph{et~al.}, ``Survey on software defect
  prediction techniques,'' \emph{International Journal of Applied Science and
  Engineering}, vol.~17, no.~4, pp. 331--344, 2020.

\bibitem{catal2009investigating}
C.~Catal and B.~Diri, ``Investigating the effect of dataset size, metrics sets,
  and feature selection techniques on software fault prediction problem,''
  \emph{Information Sciences}, vol. 179, no.~8, pp. 1040--1058, 2009.

\bibitem{wohlin2012experimentation}
C.~Wohlin, P.~Runeson, M.~H{\"o}st, M.~Ohlsson, B.~Regnell, and A.~Wessl{\'e}n,
  \emph{Experimentation in Software Engineering}, ser. Computer Science.\hskip
  1em plus 0.5em minus 0.4em\relax Springer Berlin Heidelberg, 2012.

\bibitem{yin2009case}
R.~Yin, \emph{Case Study Research: Design and Methods}, ser. Applied Social
  Research Methods.\hskip 1em plus 0.5em minus 0.4em\relax SAGE Publications,
  2009.

\bibitem{tantithamthavorn2016towards}
C.~Tantithamthavorn, ``Towards a better understanding of the impact of
  experimental components on defect prediction modelling,'' in \emph{2016
  IEEE/ACM 38th International Conference on Software Engineering Companion
  (ICSE-C)}.\hskip 1em plus 0.5em minus 0.4em\relax IEEE, 2016, pp. 867--870.

\bibitem{ali2023hyperparameter}
Y.~A. Ali, E.~M. Awwad, M.~Al-Razgan, and A.~Maarouf, ``Hyperparameter search
  for machine learning algorithms for optimizing the computational
  complexity,'' \emph{Processes}, vol.~11, no.~2, p. 349, 2023.

\bibitem{26c990771bd645428c33ea107259ceb5}
R.~J. Wieringa and M.~Daneva, ``Six strategies for generalizing software
  engineering theories,'' \emph{Science of computer programming}, vol. 101, pp.
  136--152, 4 2015.

\bibitem{stradowski2023industrial}
S.~Stradowski and L.~Madeyski, ``Industrial applications of software defect
  prediction using machine learning: A business-driven systematic literature
  review,'' \emph{Information and Software Technology}, vol. 159, p. 107192,
  2023.

\bibitem{gao2025current}
C.~Gao, X.~Hu, S.~Gao, X.~Xia, and Z.~Jin, ``The current challenges of software
  engineering in the era of large language models,'' \emph{ACM Transactions on
  Software Engineering and Methodology}, vol.~34, no.~5, pp. 1--30, 2025.

\end{thebibliography}

\end{document}